\documentclass{aa}
\usepackage{graphicx}
\usepackage{txfonts}
\usepackage{natbib}

\newcommand{\kms}{km~s$^{-1}$}

\newcommand{\degree}{\ensuremath{^\circ}}

\begin{document}

\title{Temporal evolution of the Evershed flow in sunspots\thanks{Appendices are
    only available in electronic form at \newline {\tt http://www.edpsciences.org}}}
\subtitle{II. Physical properties and nature of Evershed clouds}

\author{D.\ Cabrera Solana\inst{1}, L.R.\ Bellot Rubio\inst{1}, J.M.\
Borrero\inst{2}, and J.C.\ del Toro Iniesta\inst{1} } 
\institute{Instituto de Astrof\'{\i}sica de Andaluc\'{\i}a, CSIC, Apdo.\ 
3004, 18080 Granada, Spain; [dcabera,lbellot,jti]@iaa.es
\and High Altitude Observatory, NCAR, 3080 Center Green Dr.\ CG-1, 80301,
Boulder CO, USA; borrero@ucar.edu}

\offprints{L.R.\ Bellot Rubio}
\date{Received 08 May 2007 / Accepted 04 September 2007}

\abstract {Evershed clouds (ECs) represent the most conspicuous variation 
of the Evershed flow in sunspot penumbrae.} 
{We determine the physical properties of ECs from high spatial and temporal
resolution spectropolarimetric measurements. }
{The Stokes profiles of four visible and three infrared spectral lines are
subject to inversions based on simple one-component models as well as more
sophisticated realizations of penumbral flux tubes embedded in a static
ambient field (uncombed models).}  {According to the one-component inversions,
the EC phenomenon can be understood as a perturbation of the magnetic and 
dynamic configuration of the penumbral filaments along which these structures 
move. The uncombed inversions, on the other hand, suggest that ECs are the 
result of enhancements in the visibility of penumbral flux tubes. We
conjecture that the enhancements are caused by a perturbation of the
thermodynamic properties of the tubes, rather than by changes in the vector
magnetic field. The feasibility of this mechanism is investigated performing 
numerical experiments of thick penumbral tubes in mechanical equilibrium 
with a background field.}  
{While the one-component inversions confirm many of the
properties indicated by a simple line parameter analysis (Paper I of this
series), we tend to give more credit to the results of the uncombed inversions
because they take into account, at least in an approximate manner, the fine
structure of the penumbra.}

\keywords{Sunspots -- Sun: magnetic fields -- Sun: photosphere}

\titlerunning{Physical properties and nature of Evershed clouds}
\authorrunning{Cabrera Solana et al.}
\maketitle

\section{Introduction}

Evershed clouds (ECs) are patches of enhanced Doppler signals that move in
the penumbra of sunspots. Surprisingly, there exist very few measurements of
their magnetic fields. \citet{1994ApJ...430..413S} followed the evolution of
ECs using Dopplergrams and longitudinal magnetograms taken at the Swedish
Vacuum Telescope on La Palma. They observed cases of ECs with weak magnetogram
signals, but this result could not be interpreted unambiguously because the
magnetograms were sensitive to both field strength and field inclination
variations.

\citet[][hereafter Paper I]{cabrera.2007_paper1} used simple parameters
derived from visible and infrared lines to characterize the polarimetric
properties of ECs. While a complete observational description was possible,
the physical mechanism behind the EC phenomenon could not be studied without 
a more involved analysis of the measurements.

Here we determine the physical properties of ECs by means of Stokes
inversions, in an attempt to shed light on their nature. Two different models
are employed: one-component atmospheres, and uncombed models that take into
account the fine structure of the penumbra
(Sect.~\ref{sec:inv_proc}). Sections \ref{one_component} and \ref{uncombed}
describe the physical properties of the ECs and the penumbral filaments
hosting them, as well as their variation across the penumbra. Based on these
results, in Sect.~\ref{discussion} we examine different mechanisms that could
explain the EC phenomenon.

\section{Observations}
For a complete description of the data sets the reader is referred to Paper~I.
Time sequences of a small portion of the center-side penumbra of AR 10781 were
taken with the two polarimeters of the German Vacuum Tower Telescope (VTT) on
Tenerife. The polarimeters were operated simultaneously to record the Stokes
parameters of four lines at 630~nm and three lines at 1565~nm.  The cadence of
the scans was 3.9 min on June 30, 2005 and 2 min on July 1, 2005, when
the spot was located at heliocentric angles of 43$^\circ$ and 35$^\circ$,
respectively. Fifteen ECs moving from the mid to the outer penumbra were
identified as structures of enhanced Doppler velocities during the 236 min
covered by the scans. Thanks to the adaptive optics system of the VTT, the
spatial resolution of the infrared measurements ($\sim$0\farcs6) is one of the
highest ever reached in ground-based spectropolarimetry.

\section{Stokes inversions}
\label{sec:inv_proc}

\subsection{One-component models}

To invert the observed Stokes profiles we first use one-component models,
i.e., we assume the whole resolution element to be occupied by a single,
laterally homogeneous atmosphere. It is believed that penumbrae are formed by
at least two magnetic components with different field inclinations
\citep{1993ApJ...403..780T, 1993A&A...275..283S, 2000A&A...361..734M,
2004A&A...427..319B, 2005A&A...434..317B, 2005A&A...436.1087L,2005A&A...436..333B,
2007astro.ph..3021S}, so the physical quantities retrieved from these
inversions must be understood as an average of the various magnetic
atmospheres coexisting in the resolution element \citep{bellot.2003}. The
simplicity of the model may lead to erroneous conclusions, but it is
instructive to perform such inversions to allow comparisons with earlier
works.

The temperature stratification of the atmosphere is modified with 3
nodes\footnote{Nodes are grid points of the model atmosphere where
perturbations are sought. The new model atmosphere is constructed by
interpolating the perturbations at the nodes to all grid points.}. As
demonstrated in Paper I, visible and IR lines display different Doppler
velocities. Hence, we allow for a gradient of LOS velocity (i.e., 2 nodes) to
fit all the lines simultaneously. In contrast, the magnetic field is
supposed to be constant along the atmosphere.  The inversion also returns
height-independent macro- and micro-turbulent velocities, modeling the
presence of unresolved motions. As the amount of stray light contamination is
expected to depend on wavelength, the inversion code determines this
contribution separately for each spectral range, which adds two more free
parameters. In this way, the Stokes profiles emerging from the pixel are
computed as:
\begin{equation}
\vec{I} = (1 - \alpha) \vec{I}_{\rm m}  + \alpha \vec{I}_{\rm stray},
\label{eq:inv_1c_pen}
\end{equation}
where $\alpha$ is the stray light factor, $\vec{I}_{\rm stray}$ the average
quiet Sun intensity profile representing the stray light contribution, and
$\vec{I}_{\rm m}$ the Stokes vector emerging from the model atmosphere. 
Our one-component models and those of \citet{2004A&A...422.1093B} are 
different, since we consider height-independent magnetic fields. 

\begin{figure}
\begin{center}
\scalebox{0.3}{\includegraphics[bb=0 30 533 430,clip]{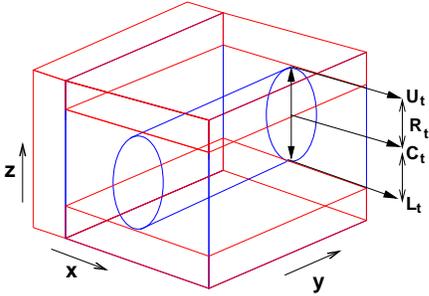}}
\vspace*{.5em}
\caption{Illustration of the atmospheres filling the resolution element in an 
uncombed penumbra. The areas enclosed by the red and blue lines represent the 
background and tube atmospheres, respectively. The embedded tube is characterized by its
center position ($C_{\rm t}$) and radius ($R_{\rm t}$). $U_{\rm t}$ and
$L_{\rm t}$ are the upper and lower boundaries of the tube. The $y$-axis
is the sunspot's radial direction and $z$ the local vertical. \label{fig:geom_mod}}
\end{center}
\end{figure}

The inversions are carried out using the SIR code
\citep{1992ApJ...398..375R} with a total of 12 free parameters.
 
\subsection{Uncombed models}
The typical size of the fine structure of the penumbra is 150-250~km or even
smaller. At the spatial resolution of our observations, atmospheres with very
different properties are likely to be mixed in the resolution element. These
atmospheres must be interlaced not only in the horizontal but also in the
vertical direction to explain the non-zero Stokes $V$ area asymmetries
observed in visible and infrared lines. The uncombed model proposed by
\cite{1993A&A...275..283S} incorporates the interlacing of field lines in a
natural way. This model envisages the penumbra as a collection of horizontal
flux tubes embedded in a more vertical background field, which agrees with the
results of spectropolarimetric analyses \citep[e.g.,][]{2001ApJ...549L.139D,
2002A&A...381..668S,2004A&A...427..319B, 2004A&A...422.1093B,
2005A&A...436..333B,2006A&A...450..383B,beck.thesis}.

To account for the uncombed structure of the penumbra, at least to first
order, the horizontal and vertical interlacing of the flux tubes and the
background must be considered. To this end, we use a two-component
model in which a fraction of the pixel is assumed to be filled by a background
atmosphere and the rest is occupied by the background and tube atmospheres
interlaced in the vertical direction (cf.\ Fig.~\ref{fig:geom_mod}).  The
emergent Stokes profiles are then the combination of the background,
$\vec{I}_{\rm b}$, and tube, $\vec{I}_{\rm t}$, profiles:
\begin{equation}
\vec{I} = (1-\alpha) [f \vec{I}_{\rm t}  + (1-f) \vec{I}_{\rm b}]+ \alpha \vec{I}_{\rm stray}.
\label{eq:inv_2c}
\end{equation}
The filling factor $f$ represents the fractional area of the pixel
occupied by the flux tube.

\begin{figure}
\begin{center}
\scalebox{0.45}{\includegraphics{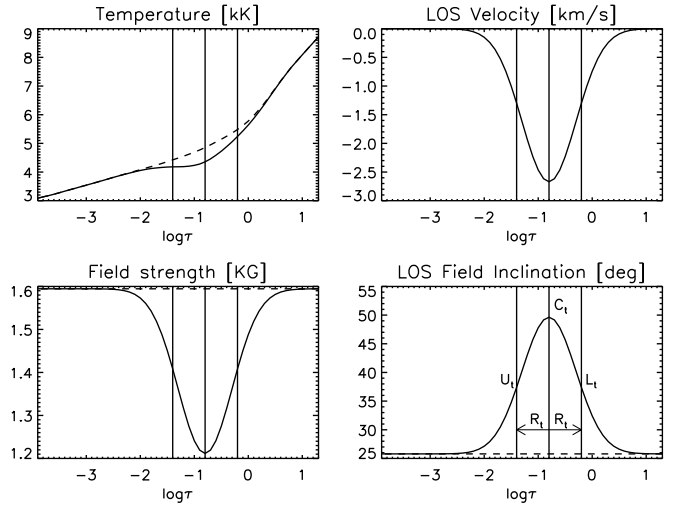}}
\caption{Example of an initial uncombed model consisting of the background
(dashed) and tube (solid) atmospheres. The gaussian is initially placed at
$\log{\tau} = -0.8$ and has a half-width of $\Delta\log{\tau} =
0.5$. \label{fig:gauss_mod}}
\end{center}
\end{figure}

The tube atmosphere is constructed adding gaussian perturbations to the
stratifications of the background component (see Fig.~\ref{fig:gauss_mod}).
The half-width ($R_{\rm t}$) and position ($C_{\rm t}$) of the gaussian are
free parameters derived from the inversion and are the same for all the
atmospheric parameters.  In contrast, the amplitudes of the gaussian depend on
the parameter.  For obvious reasons, we identify $R_{\rm t}$ and $C_{\rm t}$
with the radius and central position of the tubes. Except for the temperature,
the atmospheric quantities of the background are assumed to be constant with
optical depth. The results of \cite{2001ApJ...549L.139D},
\cite{2004A&A...427..319B}, and \cite{2004A&A...422.1093B} demonstrate that
the background atmosphere harbors much smaller velocities than the tube
component, so we take the background to be at rest.

The uncombed inversions have been performed with the SIRGAUS code
\citep{bellot.2003}. This code does not account for different stray
light contaminations in the different spectral ranges, but systematic errors
in the determination of the stray light should produce little changes of the
magnetic field properties. The number of free parameters is 17 (Table
\ref{tab:inv_2c}).

\begin{table}
\tabcolsep .6em
\caption{Number of nodes employed in the uncombed inversion. $-1$ 
indicates that the two atmospheric components have the same value 
of the corresponding parameter.
\label{tab:inv_2c}}
\begin{tabular}{l c c} 
\hline
\hline
\multicolumn{1}{l}{Parameter} &  
\multicolumn{1}{c}{Background} & \multicolumn{1}{c}{Tube} \\ \hline  	
temperature        & 2  &  1   \\
velocity           & 0  &  1   \\
field strength     & 1  &  1    \\
inclination        & 1  &  1      \\
azimuth            & 1  &  1     \\
microturbulence    & 1  &  1     \\
macroturbulence    & $-1$ &  1   \\
stray light        & $-1$ &  1   \\
filling factor     & $-1$ &  1   \\
gaussian position  &      & 1   \\
gaussian width     &      & 1   \\
\hline  		    
\end{tabular}
\end{table}

\subsection{Quality of the fits}
\label{sec:fits_2c}

\begin{figure}
\begin{center}
\scalebox{0.28}{\includegraphics[bb= -128 -55 740 850,clip]{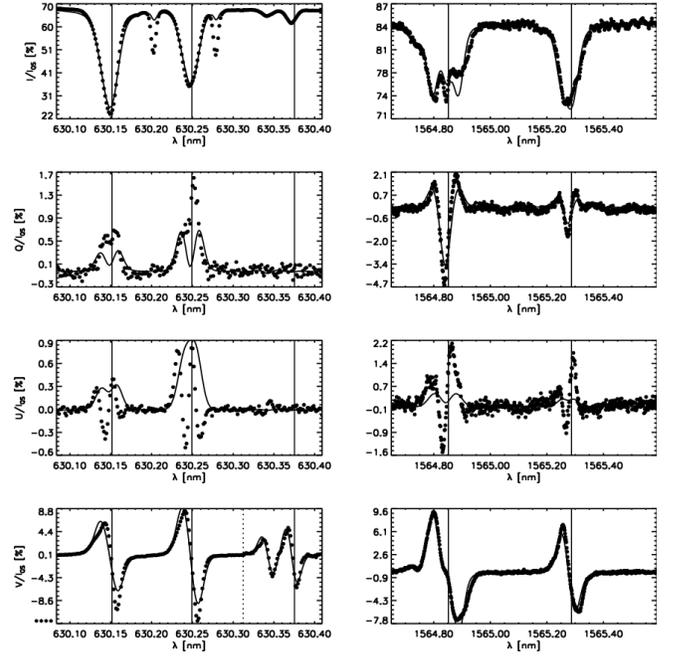}}
\caption{Example of Stokes profiles observed at $r = 0.3$ (dots) and
best-fit profiles resulting from the one-component inversions (solid
lines). The ${\rm O_2}$ teluric blends present in the visible spectra were not
inverted, but they appear in the best-fit profiles due to stray light
contamination. The vertical solid lines indicate the rest wavelengths of the
lines. The Stokes $V$ signal around \ion{Ti}{i} 630.38~nm has been multiplied
by six for better visibility (to the right of the vertical dotted
line). \vspace*{-1.8em}
\label{fig:30inv1c_profiles}}
\end{center}
\end{figure}

As an example, Fig.~\ref{fig:30inv1c_profiles} shows the best-fit profiles
resulting from the one-component inversion of a pixel located at a normalized
radial distance $r = 0.3$ ($r$ is defined such that 0 represents the inner
penumbral boundary and 1 the edge of the spot). The observed profiles
are rather asymmetric, with Stokes $U$ exhibiting four lobes. Taking into
account the simplicity of the model, the match between observed and synthetic
profiles is reasonably good in both spectral ranges. This suggests that the
scenario adopted for the one-component inversions is appropriate to determine
the 'average' properties of the atmosphere.

\begin{figure}
\begin{center}
\scalebox{0.28}{\includegraphics[bb= -128 -53 740 850,clip]{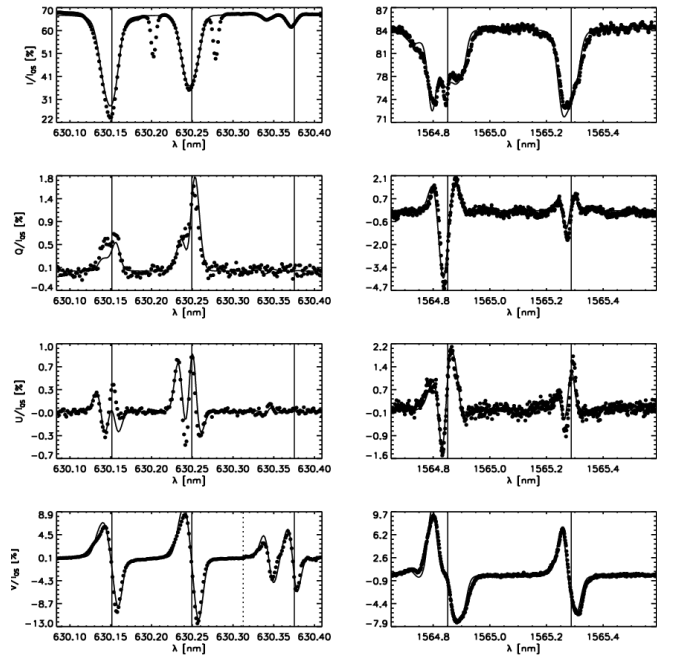}}
\caption{Same as Fig.~\ref{fig:30inv1c_profiles}, for the uncombed inversion. 
\label{fig:30invgauss_profiles_b} \vspace*{-1.5em}}
\end{center}
\end{figure}

Figure~\ref{fig:30invgauss_profiles_b} displays the same observed spectra and
the best-fit profiles resulting from the uncombed inversion. The Stokes $V$
area asymmetries and the shapes of Stokes $Q$ and $U$ are now reproduced in
much greater detail. The use of the uncombed model considerably improves the 
quality of the fits as a result of its more realistic assumptions.

\section{One-component view of the EC phenomenon}
\label{one_component}
\subsection{Physical properties of ECs}
\label{sec:phys_1c}

\begin{figure*}
\begin{center}
\scalebox{0.67}{\includegraphics[bb=54 435 780 
1005,clip]{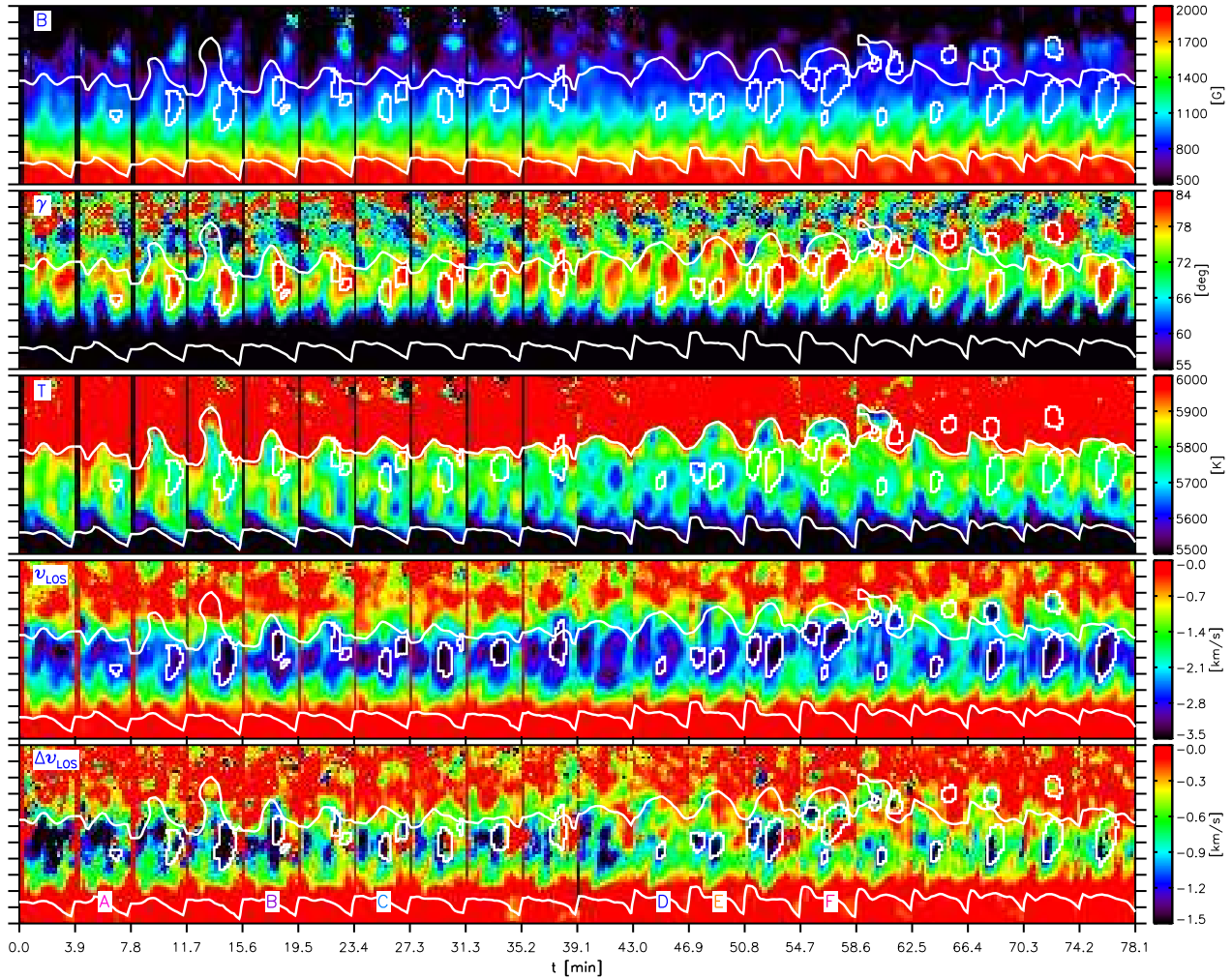}}
\caption{{\em Top} to {\em bottom}: Magnetic field strength,
field inclination, temperature at $\log \tau =0$, LOS velocity 
at $\log \tau =-0.5$, and differences between the unsigned
LOS velocities at $\log \tau =0$ and $\log \tau =-0.5$. White contours
outline the ECs and the inner and outer penumbral boundaries. The
letters at the bottom of the last panel label each EC. \label{fig:30inv1c_1}}
\end{center}
\end{figure*}

Figures~\ref{fig:30inv1c_1} and \ref{fig:30inv1c_3} show the atmospheric
parameters inferred from the inversion of the June 30 data set. The different
panels display the magnetic field strength, magnetic field inclination in the
local reference frame (LRF), the temperature $T$ at $\log \tau =0$, the LOS
velocity $v_{\rm LOS}$ at $\log \tau =-0.5$, and the difference between the
unsigned LOS velocity in two different layers, $\Delta v_{\rm LOS} = |v_{\rm
LOS}(\log \tau =-0.5)| - |v_{\rm LOS}(\log \tau =0)|$. This last quantity is
proportional to the gradient of velocity with optical depth.

The vector magnetic field displays the typical behavior observed in sunspots:
it is weaker and more horizontal as the radial distance increases. It does not
drop to zero outside the visible border of the spot, which is the signature
of the sunspot magnetic canopy \citep{1994A&A...283..221S,
2001ApJ...547.1130W, 2006A&A...454..975R}. The ECs are visible as coherent
structures with increased LOS velocity and field inclination. The presence of
many moving magnetic features in the surroundings of the spot is obvious (see
the arrows plotted in the first panel of Fig.~\ref{fig:30inv1c_3}). Their
origin and nature will be discussed elsewhere \citep{cabrera.2008}.

For each EC, we calculate the maximum field strength, field inclination in the
LRF, and LOS velocity at $\log \tau =-0.5$ following the procedure described
in Sect.\ 6 of Paper I. Assuming that the velocity and magnetic field vectors
are aligned, the modulus of the velocity vector ({\em the flow velocity}) is
simply $|v|=v_{\rm LOS}/\cos{ \gamma_{\rm LOS} }$. We also determine the
maximum difference between the flow velocities at $\log \tau = -0.5$ and $\log
\tau = 0$ ($\Delta |v|$). The results are summarized in
Table~\ref{tab:ecs_inv1c}.

The maximum field strengths attained by the ECs range from 930~G to 1070~G
with an average of 1005~G on June 30, and from 1300~G to 1500~G with an
average of 1408~G on July 1. In Paper~I we discussed the compatibility of the
linear-to-circular polarization ratios and LOS velocities with an scenario
where the magnetic and velocity fields of the ECs are nearly horizontal to the
solar surface. Table~\ref{tab:ecs_inv1c} confirms the existence of very
inclined magnetic fields in the ECs: the average field inclination is deduced
to be of the order of 82$^\circ$ for the June 30 observations and 71$^\circ$
for the July 1 data set.

Both $|v|$ and $\Delta |v|$ display higher values on June 30, i.e.\ when 
the spot is farther from disk center. The values of $|v|$ are around
$4.5$~\kms~and $2.6$~\kms~on June 30 and July 1, while $\Delta |v|$ is about
$-0.8$~\kms and $-0.3$~\kms on June 30 and July 1, respectively. The changes
in $|v|$ and $\Delta |v|$ cannot be explained by LOS effects. Thus, the
properties of the spot seem to have varied between the two observations.

\begin{table}
\tabcolsep .9em
\caption{Physical properties of the observed ECs: field strength, 
inclination in the LRF, flow velocity at $\log \tau =-0.5$, and the
difference between the flow velocities at $\log \tau =-0.5$ and 
$\log \tau =0$. Errors represent the rms fluctuations of
the physical parameters within each EC. \label{tab:ecs_inv1c}}
\begin{tabular}{l c c c c} 
\hline \hline
\multicolumn{1}{l}{EC} & \multicolumn{1}{c}{$B$} & \multicolumn{1}{c}{$\gamma$} & 
\multicolumn{1}{c}{$|v|$} & \multicolumn{1}{c}{$\Delta |v|$}\\
 & [G] & [deg] & [km/s] & [km/s]  \\ \hline  
A & $1020\pm20$   & $81\pm3$ & $4.5\pm0.4$ & $-1.0\pm0.6$     \\ 
B & $1010\pm20$   & $83\pm2$ & $5.0\pm0.5$ & $-0.6\pm0.4$        \\ 
C & $1050\pm60$   & $80\pm3$ & $4.5\pm0.3$ & $-0.8\pm0.6$        \\ 
D & $1010\pm50$   & $82\pm5$ & $4.5\pm0.8$ & $-0.6\pm0.6$       \\ 
E &  $980\pm30$   & $81\pm3$ & $4.2\pm0.5$ & $-1.1\pm0.8$        \\ 
F & $1070\pm60$   & $88\pm5$ & $4.8\pm0.8$ & $-0.4\pm0.4$        \\ 
G & $1010\pm50$   & $84\pm2$ & $4.2\pm0.4$ & $-0.6\pm0.2$        \\ 
H &  $950\pm40$   & $83\pm1$ & $4.7\pm0.3$ & $-0.3\pm0.2$      \\ 
I &  $1000\pm50$   & $78\pm4$ & $4.1\pm0.3$ & $-1.0\pm0.3$       \\ 
J &  $930\pm40$   & $83\pm1$ & $4.7\pm0.3$ & $-0.9\pm0.3$       \\ 
K & $1030\pm40$   & $79\pm2$ & $4.3\pm0.3$ & $-1.3\pm0.1$       \\ 
\hline		    	          		    
Mean  & {\bf 1005} & {\bf 82} & {\bf 4.5} &  {\bf -0.8}  \\ 
\hline		    	          		    
L & $1470\pm20$  & $60\pm3$ & $1.4\pm0.5$ & $-0.2\pm0.1$       \\ 
M & $1440\pm70$   & $73\pm3$ & $2.5\pm0.5$ & $-0.2\pm0.1$        \\
N & $1430\pm110$  & $77\pm2$ & $3.1\pm0.4$ & $-0.3\pm0.1$       \\ 
O & $1290\pm30$   & $75\pm4$ & $3.2\pm0.2$ & $-0.2\pm0.1$       \\ 
\hline
Mean  & {\bf 1408} & {\bf 71} & {\bf 2.6} & {\bf -0.2}  \\ 
\hline
\end{tabular}
\end{table}

\subsection{Radial variation of the physical properties of ECs}
\label{sec:evol_rad_1c}

The variation with radial distance of the atmospheric parameters derived from
the one-component inversion is studied in this section. We calculate the
radial curves of Fig.~\ref{fig:evol_phys2} in the same way as described in
Paper~I. The comparison of the physical properties of the ECs and the penumbral 
filaments hosting them ({\em intra-spines}, according to Paper~I) shows that:

\begin{figure*}
\begin{center}
\scalebox{0.485}{\includegraphics{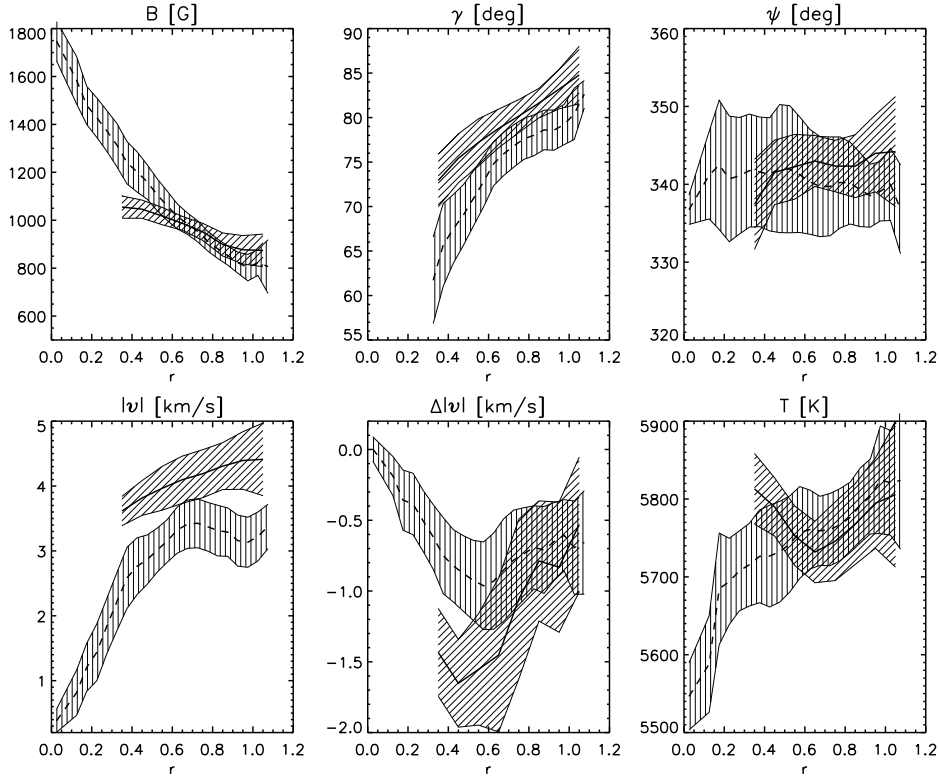}}
\caption{Radial variation of the magnetic field strength, inclination in the
LRF, azimuth, flow velocity, $\Delta |v|$, and temperature for the ECs (solid
lines) and the intra-spines along which they move (dashed lines), as inferred
from the one-component inversions. Shaded areas indicate the rms fluctuations
of the parameters. $r=0$ represents the inner penumbral boundary and $r=1$ the
outer edge of the penumbra\label{fig:evol_phys2}}
\end{center}
\end{figure*}

\begin{itemize}
\item On average, the magnetic field of the ECs is weaker than that 
of the intra-spines from the inner to the mid penumbra, but stronger in the 
outer penumbra. 

\item ECs harbor more inclined fields than the intra-spines at all radial
distances.

\item The azimuth of the field in the ECs is very similar to that found in 
the intra-spines at all radial distances, except perhaps in the outer penumbra. 

\item The flow velocity inside the ECs is greater than in the intra-spines
at all radial distances.  

\item ECs display stronger gradients of the flow velocity with optical depth 
than the intra-spines. 

\item ECs are slightly hotter than the intra-spines in the inner penumbra, 
but cooler from the mid to the outer penumbra.  

\end{itemize}

Therefore, in terms of one-component models, the magnetic, dynamic and thermal
properties of the ECs are rather different from those of the intra-spines
along which they move.

\subsection{ECs as perturbations propagating along intra-spines}
\label{sec:mod_ecs_incl}

In Paper I we found that ECs exhibit larger linear polarization signals than
the filaments hosting them. This led us to suggest that they possess more
inclined magnetic fields to the LOS. If the magnetic field and the velocity
vectors are parallel, any increase in the LOS inclination would result in
smaller LOS velocities. Thus, one has to conclude that the large Doppler
signals associated with the ECs can only be produced by an increase in the
modulus of the velocity vector, i.e., by stronger Evershed flows.

\begin{figure*}
\begin{center}
\scalebox{0.64}{\includegraphics{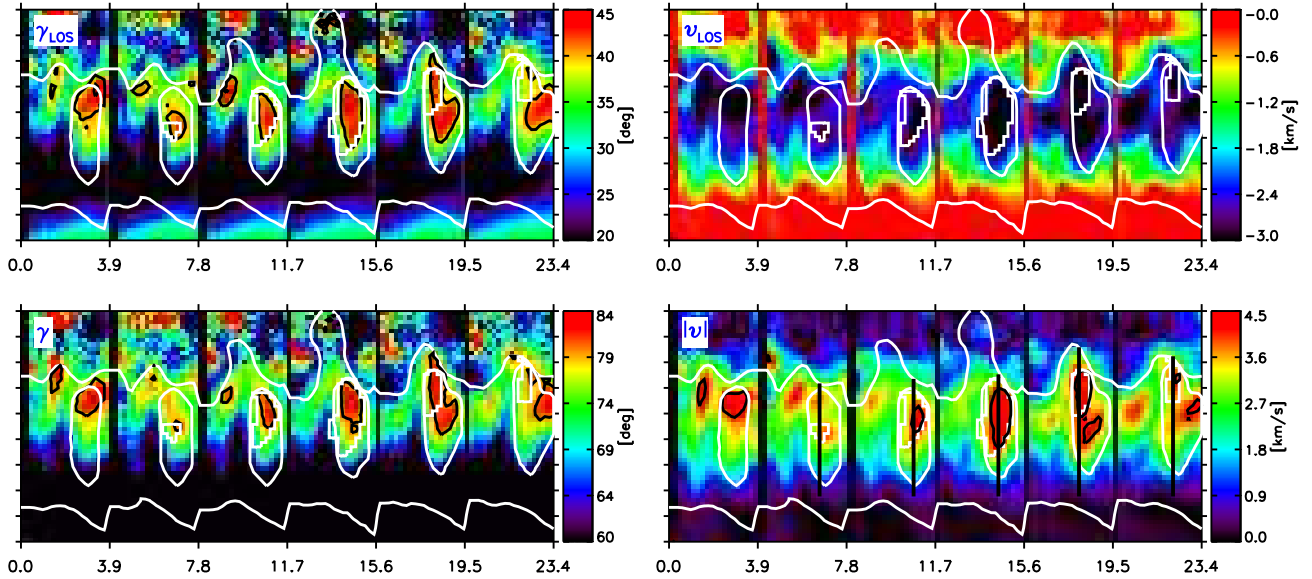}}
\caption{{\em Left:} Maps of magnetic field inclination in the LOS reference
frame ({\em top}) and the LRF ({\em bottom}). {\em Right:} Maps of LOS velocities
at $\log \tau = -0.5$ ({\em top}) and flow velocity at $\log \tau = -0.5$
({\em bottom}). White contours mark EC A, the intra-spine hosting it, and the
boundaries of the penumbra. Black contours enclose pixels with $\gamma$,
$\gamma_{\rm LOS}$, and $|v|$ greater than 80\degree, 40\degree, and
4.2~\kms\/ respectively. Each tickmark in the $y$-axis represents
1\arcsec. \label{fig:ec_filament_inv}}
\end{center}
\end{figure*}

\begin{figure*}
\begin{center}
\scalebox{0.3}{\includegraphics[bb=50 200 390 420,clip]{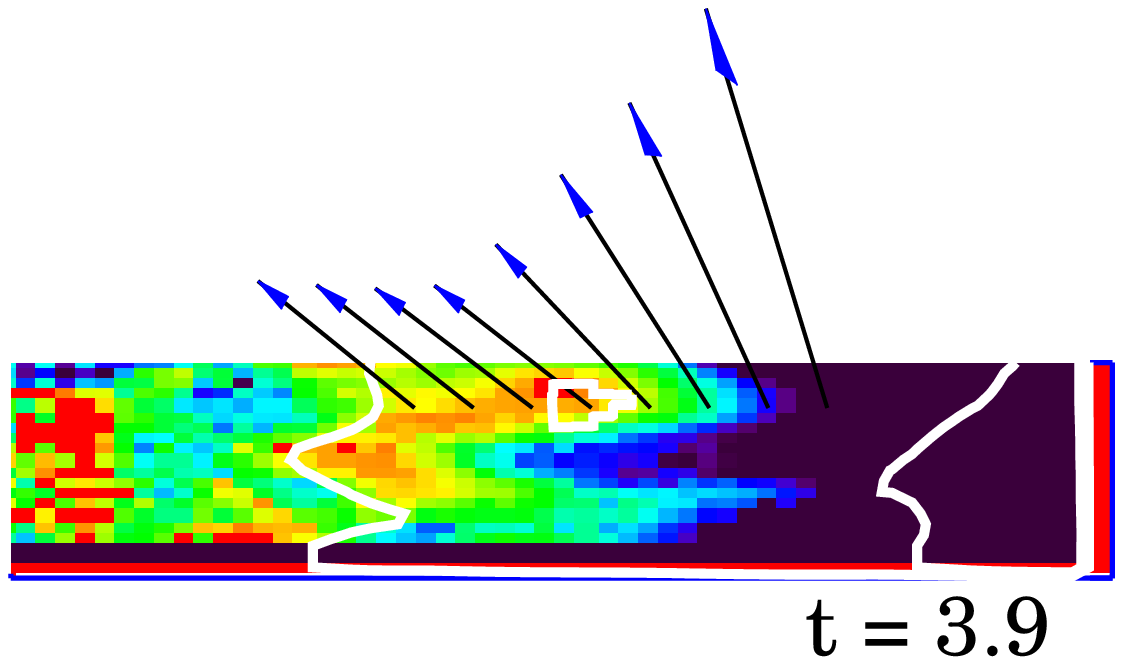}}
\scalebox{0.3}{\includegraphics[bb=50 200 390 420,clip]{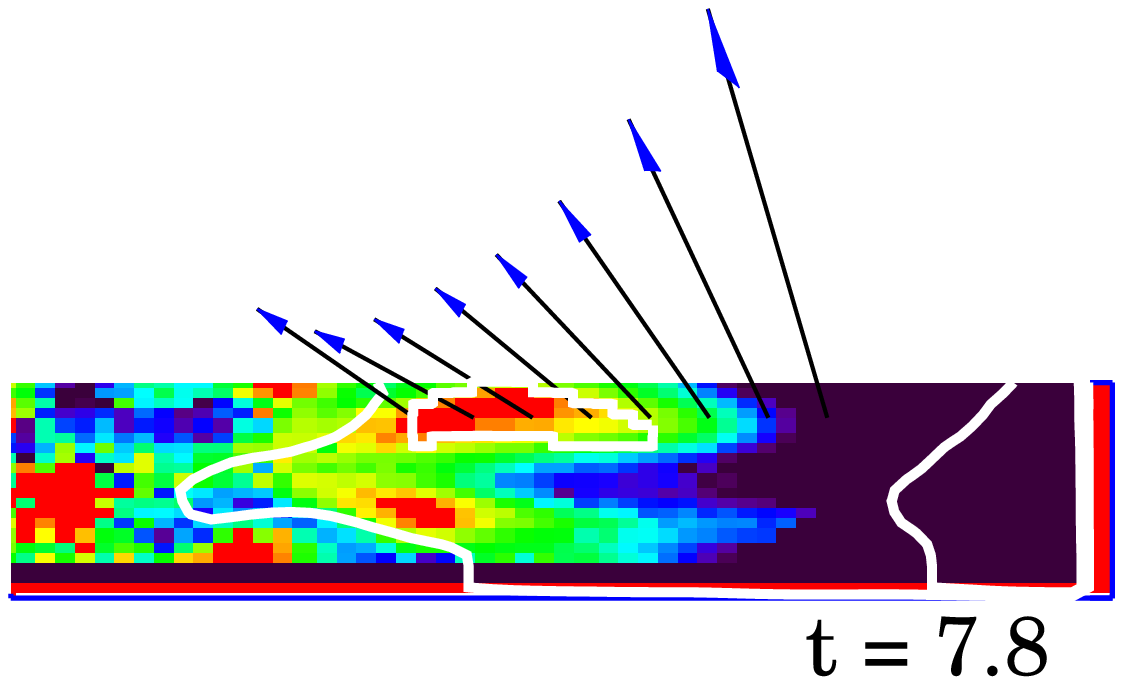}}
\scalebox{0.3}{\includegraphics[bb=50 200 390 420,clip]{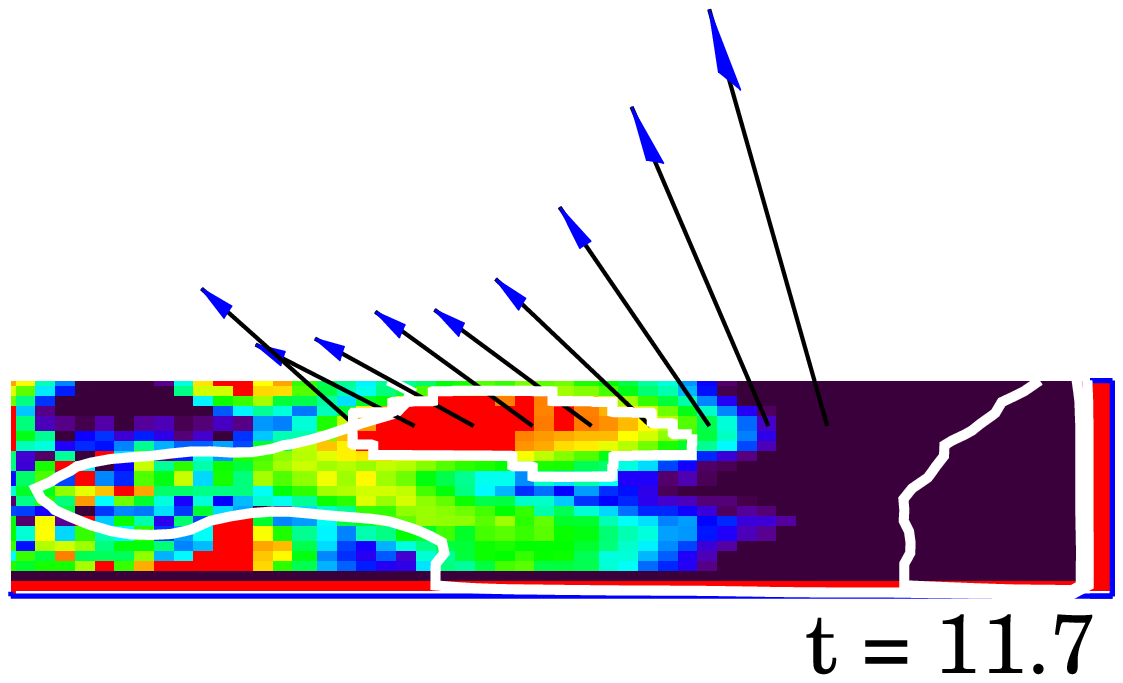}}
\scalebox{0.3}{\includegraphics[bb=50 200 390 420,clip]{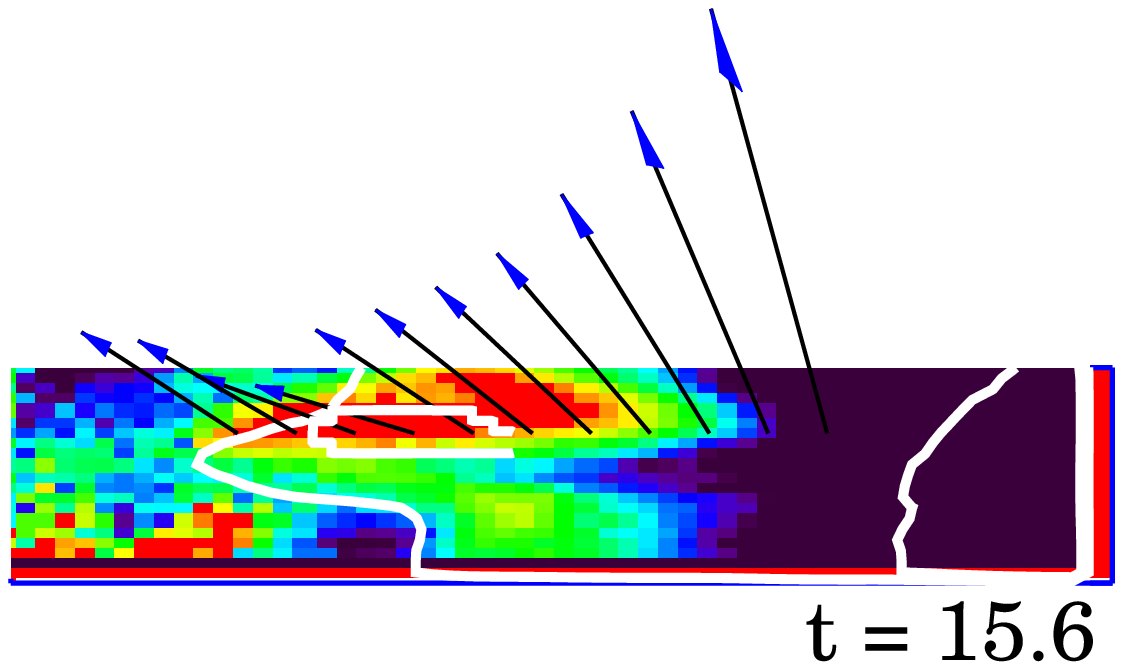}}
\scalebox{0.3}{\includegraphics[bb=50 200 390 420,clip]{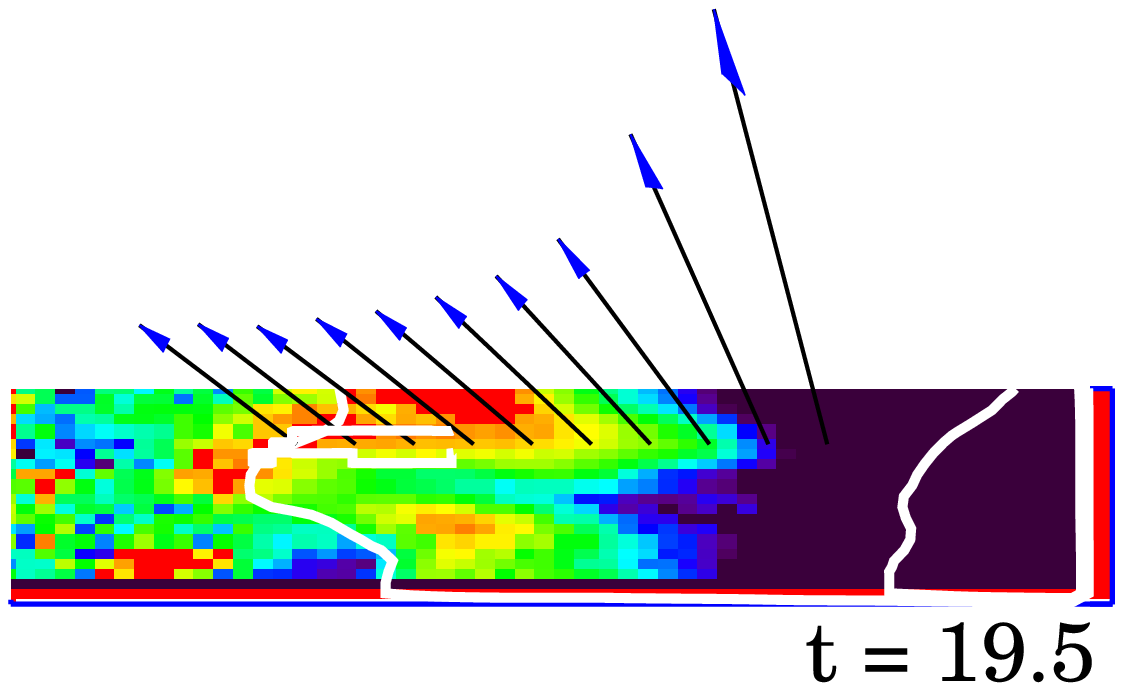}}
\caption{Variation of $\gamma$ along the cuts
shown in Fig.~\ref{fig:ec_filament_inv} as EC A propagates from the
inner ($t=3.9$~min) to the outer ($t=19.5$~min) penumbra. The color
scale ranges from 55\degree (black) to 85\degree (red). Each arrow has
the same length and indicates the orientation of the magnetic field
vector retrieved from the inversion. White contours outline EC A and
the penumbral boundaries.\label{fig:var_inc_1c}\vspace*{-1em}}
\end{center}
\end{figure*}

The maps of field inclinations and velocities displayed in
Fig.~\ref{fig:ec_filament_inv} for EC A demonstrate that this interpretation
is correct. The strong linear polarization signals seen in the EC are indeed
produced by larger field inclinations to the LOS (upper left panel), which
come from more horizontal fields (lower left panel). The higher LOS velocities
detected in the EC are the observed signatures of stronger Evershed flows
(lower right panel).
 
During their journey to the outer penumbral boundary, the ECs modify the
magnetic field of the intra-spines along which they move, making it more
horizontal locally.  After the passage of an EC, the intra-spine recovers 
its original field inclination. Figure \ref{fig:var_inc_1c} illustrates 
this behaviour for the case of EC A (the arrows indicate the orientation of the
vector magnetic field).

In view of these results, one could define the ECs as quasi-periodic
perturbations of the magnetic and velocity fields of the penumbral filaments
(intra-spines) along which they move. The main properties of these
perturbations are the following: (a) they produce enhancements of $\gamma$ and
$|v|$; (b) the amplitude of the perturbation increases as it reaches larger
radial distances; (c) when the EC starts to vanish, the amplitude of the
perturbation is reduced; and (d) the modification of $\gamma$ is perpendicular
to the direction of propagation of the perturbation (the vertical component of
the EC proper motions is small compared with the horizontal component, cf.\
Paper~I).  Except for (c), the behavior described in this section is also
observed in other type I ECs.

\section{Uncombed view of the EC phenomenon}
\label{uncombed}
The spine/intra-spine organization of the penumbra deduced from one-component
inversions can be interpreted as the azimuthal variation of the filling
factors of two components whose properties remain relatively constant at a
given radial distance \citep{2004A&A...427..319B}.  Thus, the question
naturally arises as to whether the more inclined fields and larger flow
velocities of the ECs indicated by the one-component inversions are real or an
artifact of simplistic modeling.

\begin{figure*}
\begin{center}
\scalebox{0.67}{\includegraphics[bb=54 420 780 
905,clip]{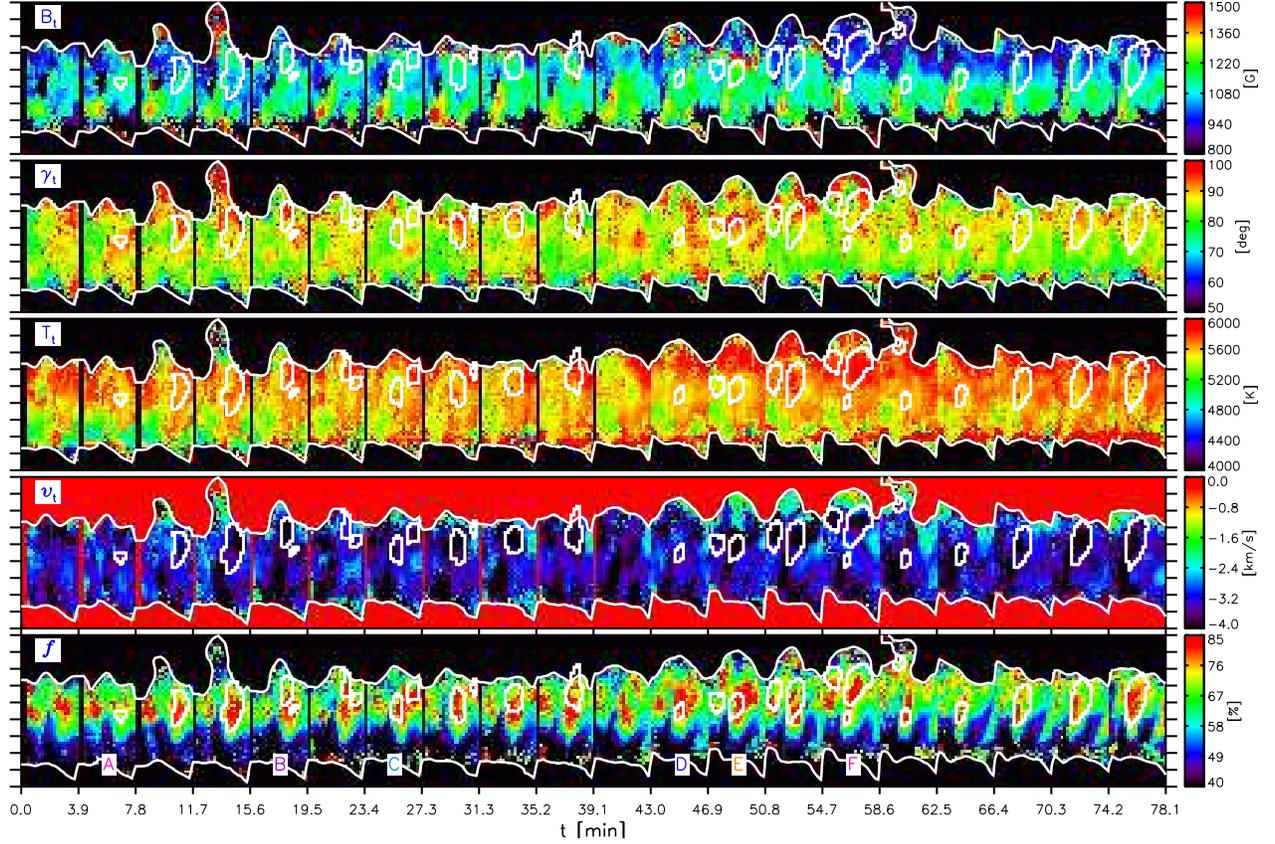}}
\caption{{\em Top} to {\em bottom}: Magnetic field strength, field
inclination, temperature at $\log \tau =0$, LOS velocity, and filling factor
of the tube atmosphere retrieved from the uncombed inversion. The parameters 
of the tubes are constructed taking the values at the tube center. White 
contours outline the ECs and the inner and outer penumbral boundaries. 
The letters at the bottom of the last panel label each EC.
\label{fig:tube1}} \end{center}
\end{figure*}

Figures~\ref{fig:tube1} and \ref{fig:tube2} show the physical properties of
the tube component derived from the uncombed inversions, for the June 30 data
set. They already make it clear that the properties of the tubes in the ECs
and the intra-spines are similar, except for the filling factor. In other
words: when the fine structure of the penumbra is considered, the ECs and the
intra-spines do not seem to have different physical properties. This is in
contrast with the results of one-component inversions
(Figs.~\ref{fig:30inv1c_1} and \ref{fig:30inv1c_3}).

\subsection{Properties of the tube and background atmospheres inside the ECs}
\label{sec:comp_mag_compo}

\begin{figure}
\begin{center}
\scalebox{0.33}{\includegraphics{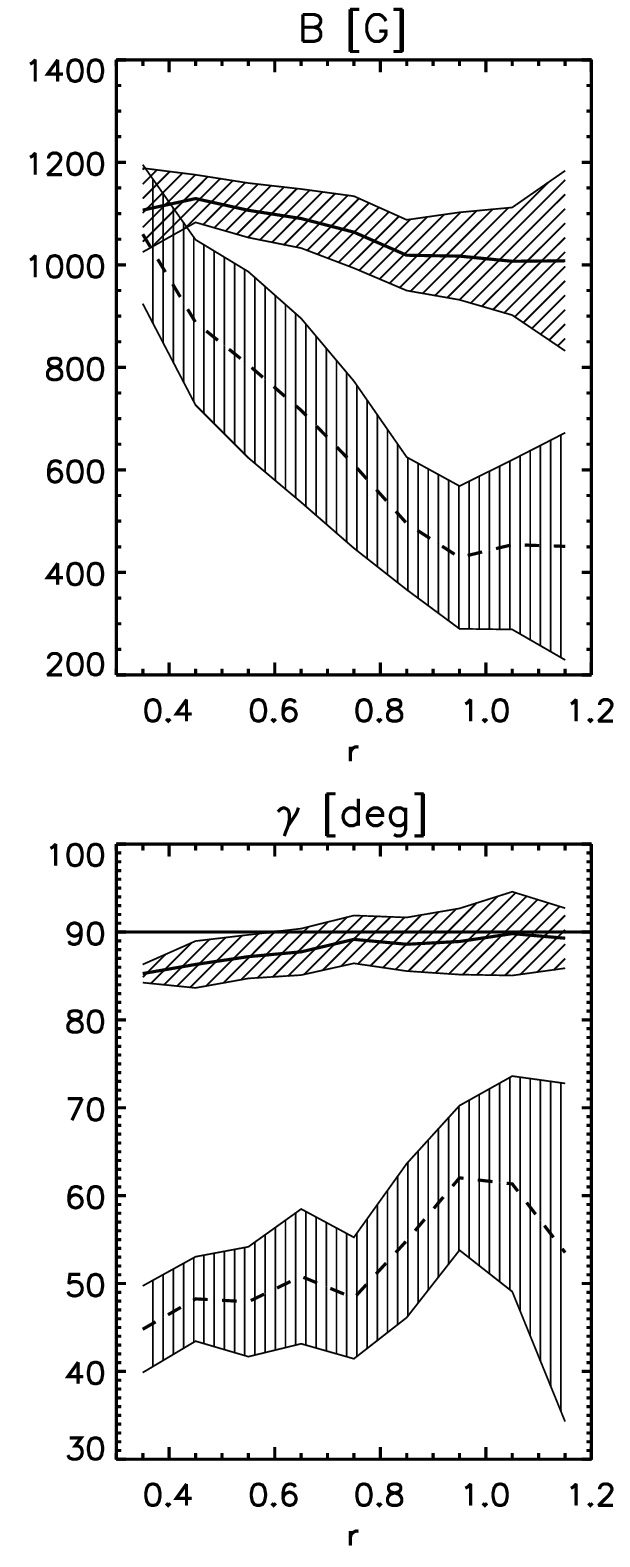}}
\scalebox{0.33}{\includegraphics{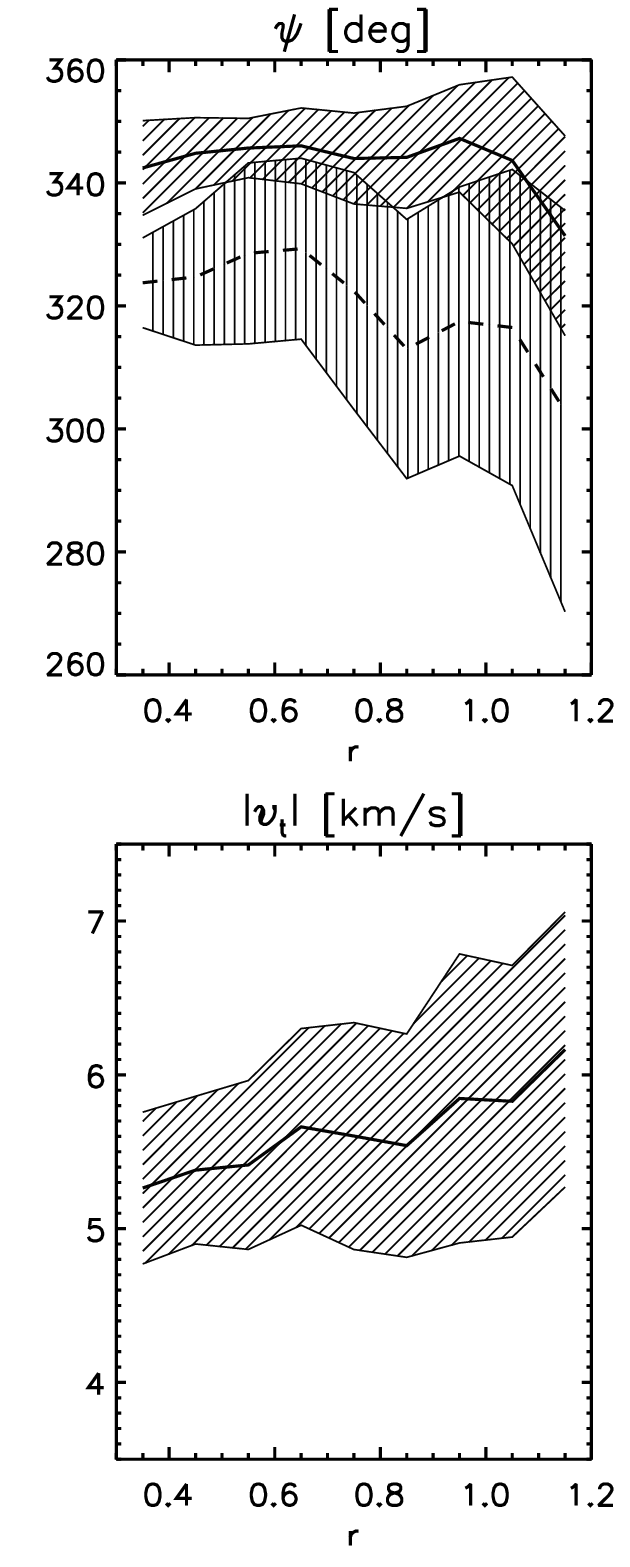}}
\scalebox{0.33}{\includegraphics{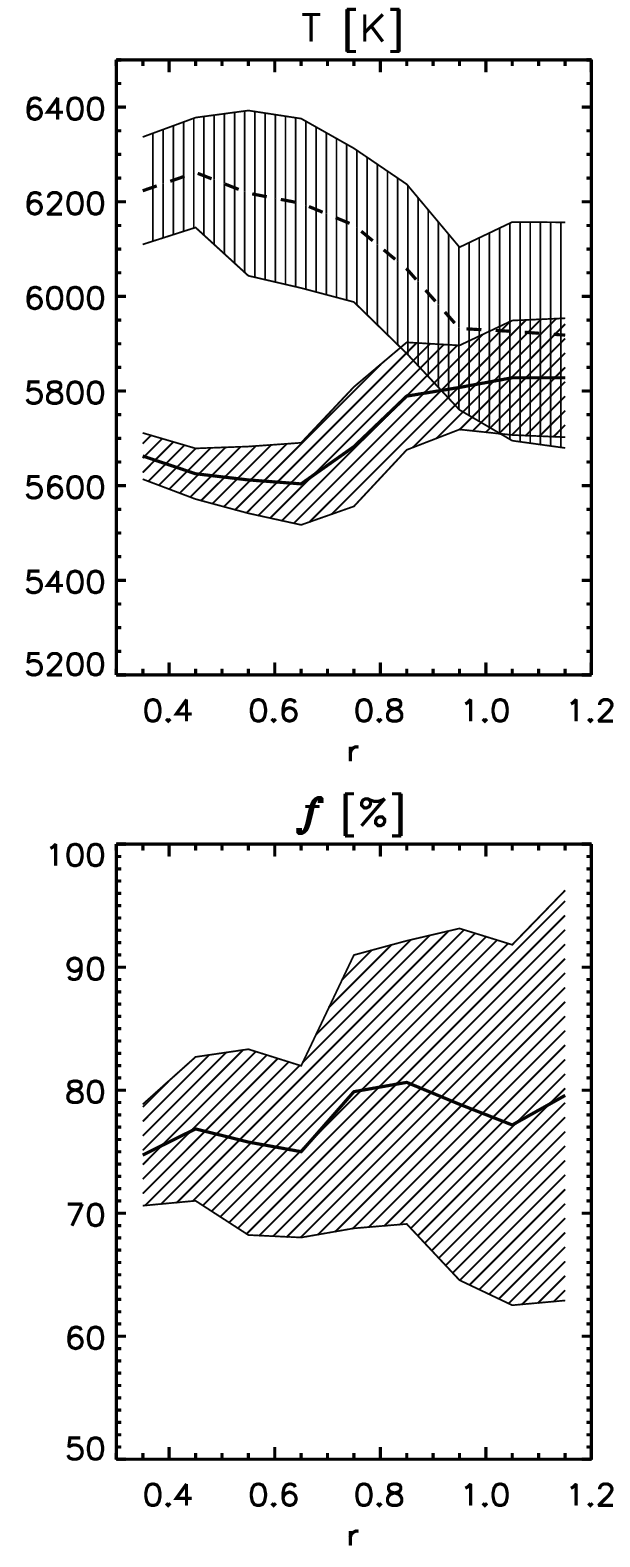}}
\caption{Radial variation of the magnetic field strength, field
inclination, field azimuth, flow velocity, temperature at $\log \tau =
0$, and filling factor of the tube (solid) and background (dashed)
components within the ECs, as derived from the uncombed inversions.
Shaded areas represent the rms fluctuations of the atmospheric
quantities at a given distance.
\label{fig:30_ecs_2c_gauss_1}\vspace*{-1.5em}} 
\end{center}
\end{figure}

Figure~\ref{fig:30_ecs_2c_gauss_1} displays the radial variation of the field
strength, LRF inclination, azimuth, flow velocity, temperature at $\log \tau =
0$, and filling factor of the tube and background atmospheres in the ECs.
Similar curves are obtained from classical two-component inversions of the
spectra \citep{cabrera.thesis}, the only difference being the weaker
background fields indicated by the uncombed inversions\footnote{These weak
fields may not be real but a a consequence of initializing the
inversion with low temperatures in the tubes.}.

As expected, the field is more inclined in the tubes than in the background
at all radial distances. In agreement with previous investigations
\citep{1997Natur.389...47W,2001ApJ...547.1130W,2000A&A...358.1122S,
2003A&A...410..695M,2004A&A...427..319B,2004A&A...422.1093B,
2005A&A...436..333B,2006A&A...450..383B}, field lines diving back to the solar
interior ($\gamma_{\rm t} > 90^{\circ}$) are found in the outer penumbra at
the position of the ECs. The tube and background fields show differences in
azimuth, but they are usually smaller than 20\degree.

Figure~\ref{fig:30_ecs_2c_gauss_1} demonstrates that the flow velocity
increases monotonically with radial distance within the tubes, varying by 
some 0.7~km~s$^{-1}$ from the inner to the outer penumbra. The tube 
temperatures at $\log \tau= 0$ are smaller than those of the background 
at all radial distances. The same behavior has been found by 
\cite{2005A&A...436..333B,2006A&A...450..383B}, but only in the outer
penumbra. The tube filling factor increases monotonically with radial
distance, in agreement with \citet{2005A&A...436..333B}.

\subsection{Position and width of the tubes}
\label{sec:pos_rad}

\begin{figure}
\begin{center}
\scalebox{0.41}{\includegraphics{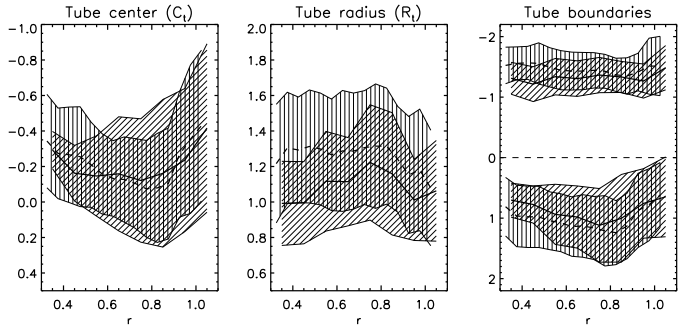}}
\caption{Center position, radius, and boundaries of the flux tubes in the ECs
(solid lines) and the intra-spines hosting them (dashed lines). The quantities
are given in units of the logarithm of the optical depth. The shaded areas
represent the standard deviations of the parameters.\label{fig:pos_width_t}}
\end{center}
\end{figure}

Figure~\ref{fig:pos_width_t} examines the center position, radius, and
boundaries of the tubes deduced from the inversion. At all radial distances,
the tube axes are located above $\log{\tau} \sim 0$, with similar values in
the ECs and the intra-spines. Within the scatter we do not detect differences
between the ECs and the intra-spines in terms of width or position of the
tube's upper boundary. The lower boundary is always below the line forming
region, so the tubes are necessarily optically thick and/or low lying. This
implies that the spectral lines only sample their upper halfs. A similar
result has been reported by \cite{2006A&A...450..383B} from uncombed
inversions of penumbral profiles.

\subsection{ECs as structures of increased filling factor and flow velocity}
\label{sec:prop_ecs_uncombed}

\begin{figure}
\begin{center}
\scalebox{0.6}{\includegraphics{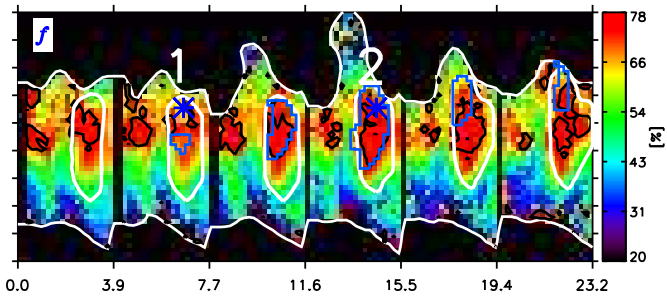}}
\caption{Maps of the filling factor of the tube ($f$) derived from the uncombed
inversions. White contours mark the intra-spine and the boundaries of the
penumbra. EC A is outlined with blue contours. Black contours delimit pixels
having $f$ larger than 77$\%$. Asterisks and numbers indicate the position of
a pixel before and during the passage of the EC.\label{fig:ec_filament_inv2c}}
\end{center}
\end{figure}

\begin{figure}
\begin{center}
\scalebox{0.44}{\includegraphics[bb= 74 370 620 623]
{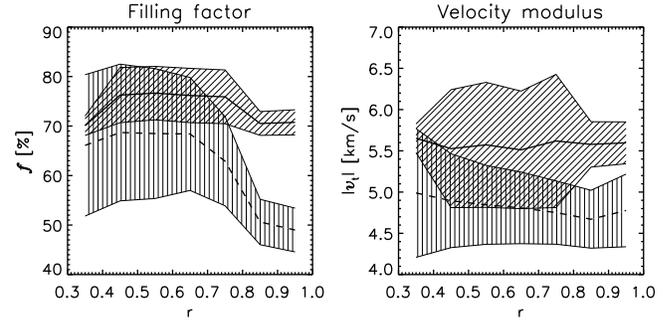}}
\caption{Radial variation of the filling factor ({\em left}) and 
flow velocity ({\em right}) inside EC A (solid lines) and its intra-spine 
(dashed lines).\label{fig:fill_modv_eca_intra}}\vspace*{-1.5em}
\end{center}
\end{figure}

Figure~\ref{fig:ec_filament_inv2c} shows maps of the filling factor of the
tube ($f$) as derived from the uncombed inversions (only the case of EC A is
considered; the other ECs behave in a similar way). The maps clearly
demonstrate that {\em ECs are regions of increased filling factors}.  The same
conclusion can be drawn from the left panel of
Fig.~\ref{fig:fill_modv_eca_intra}, where we plot the radial variation of $f$
for EC A and the rest of the intra-spine. The flow velocity is also higher in
the EC (right panel of Fig.~\ref{fig:fill_modv_eca_intra}).

\begin{figure}
\begin{center}
\scalebox{0.385}{\includegraphics{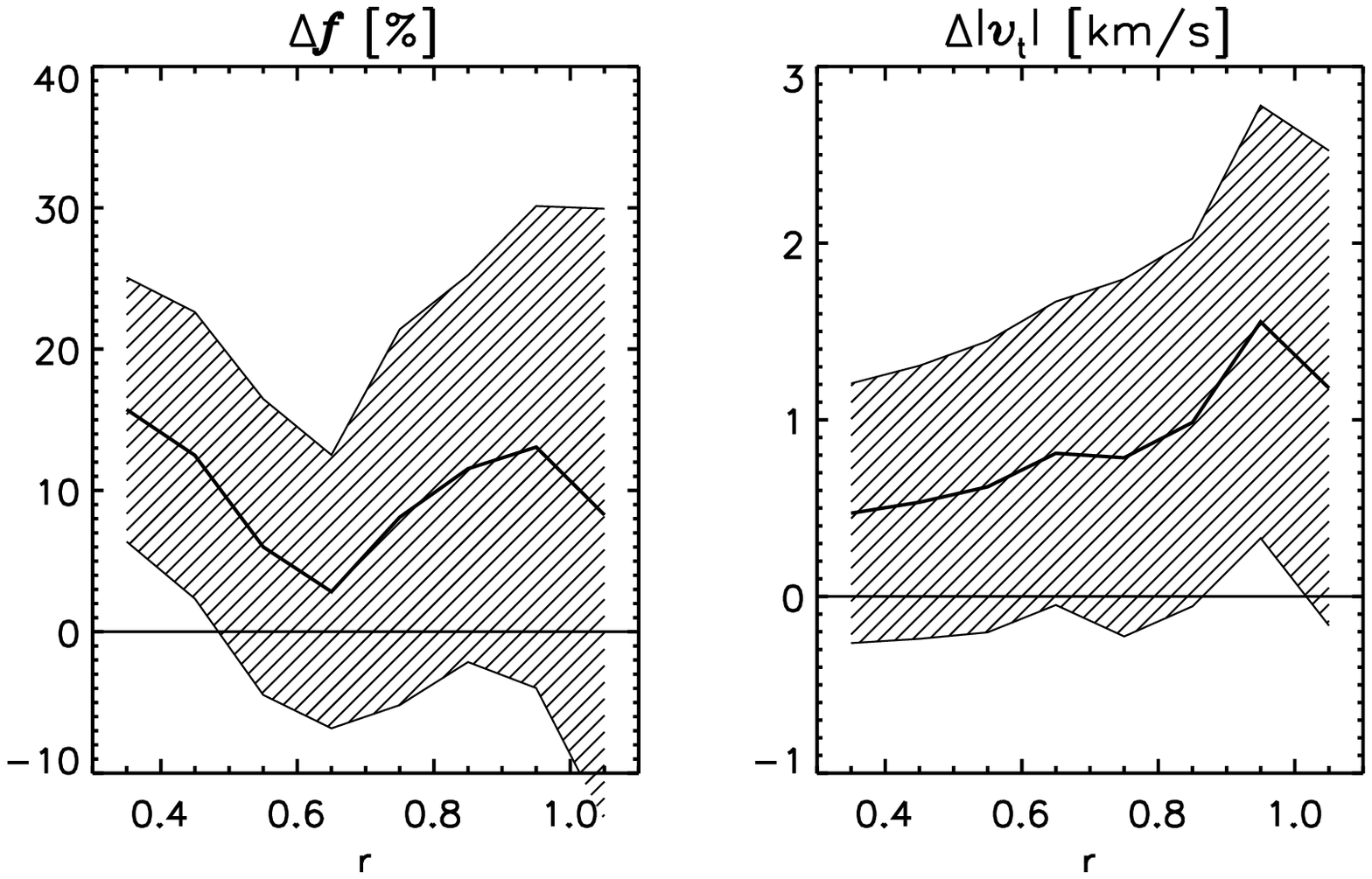}} \\
\scalebox{0.385}{\includegraphics{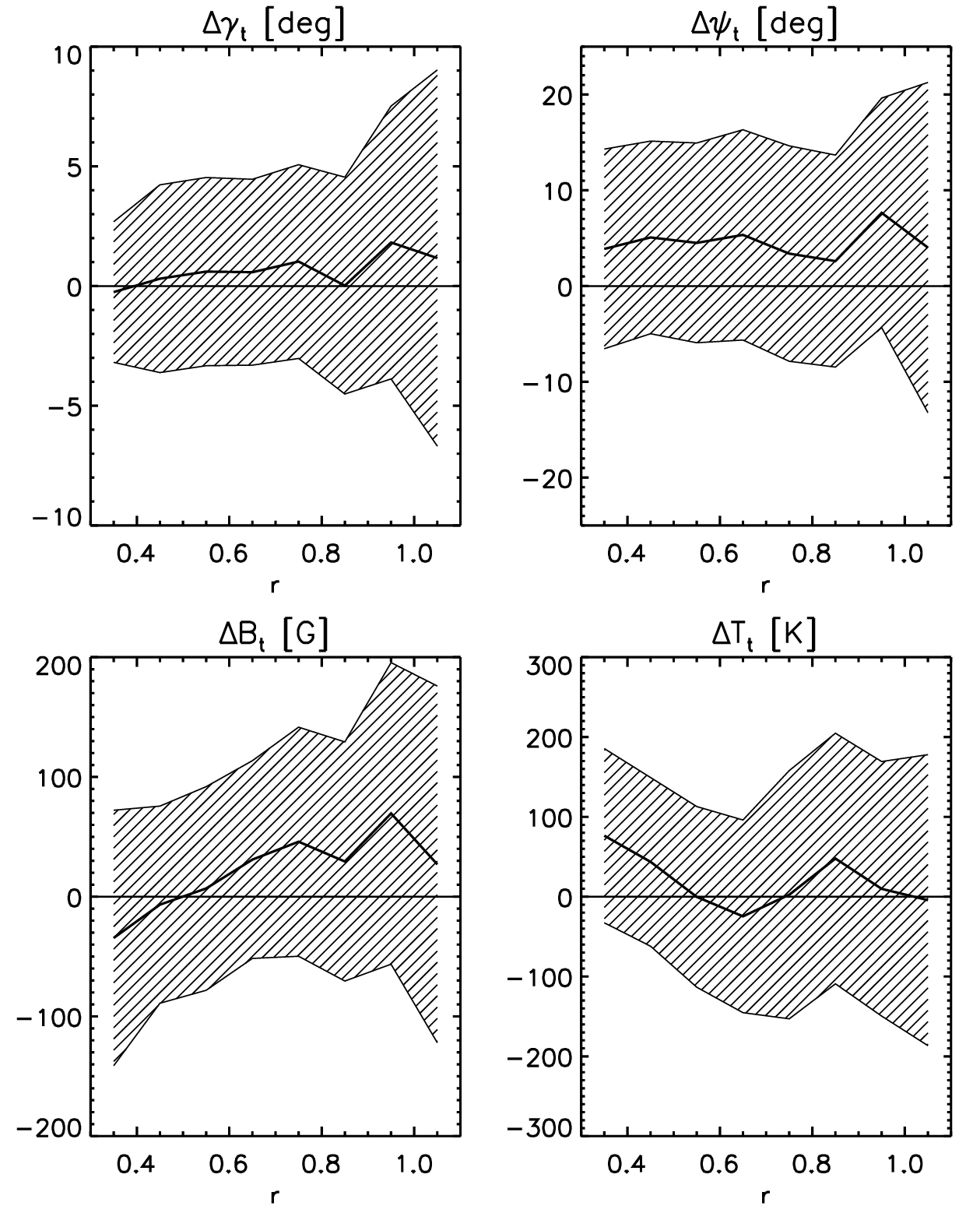}}
\caption{{\em Top:} Differences between the filling factor ({\rm
left}) and flow velocity ({\rm right}) of the tubes inside the ECs and
the intra-spines as derived from the uncombed inversions. The shaded
areas represent the rms fluctuations of the differences.  {\em
Middle:} Same, for the field strength and inclination. {\em Bottom:}
Same, for field azimuth and temperature at $\log \tau =
0$. \label{fig:dif_fill_ecs}\vspace*{-1em}}
\end{center}
\end{figure}

\begin{figure}
\begin{center}
\scalebox{0.385}{\includegraphics{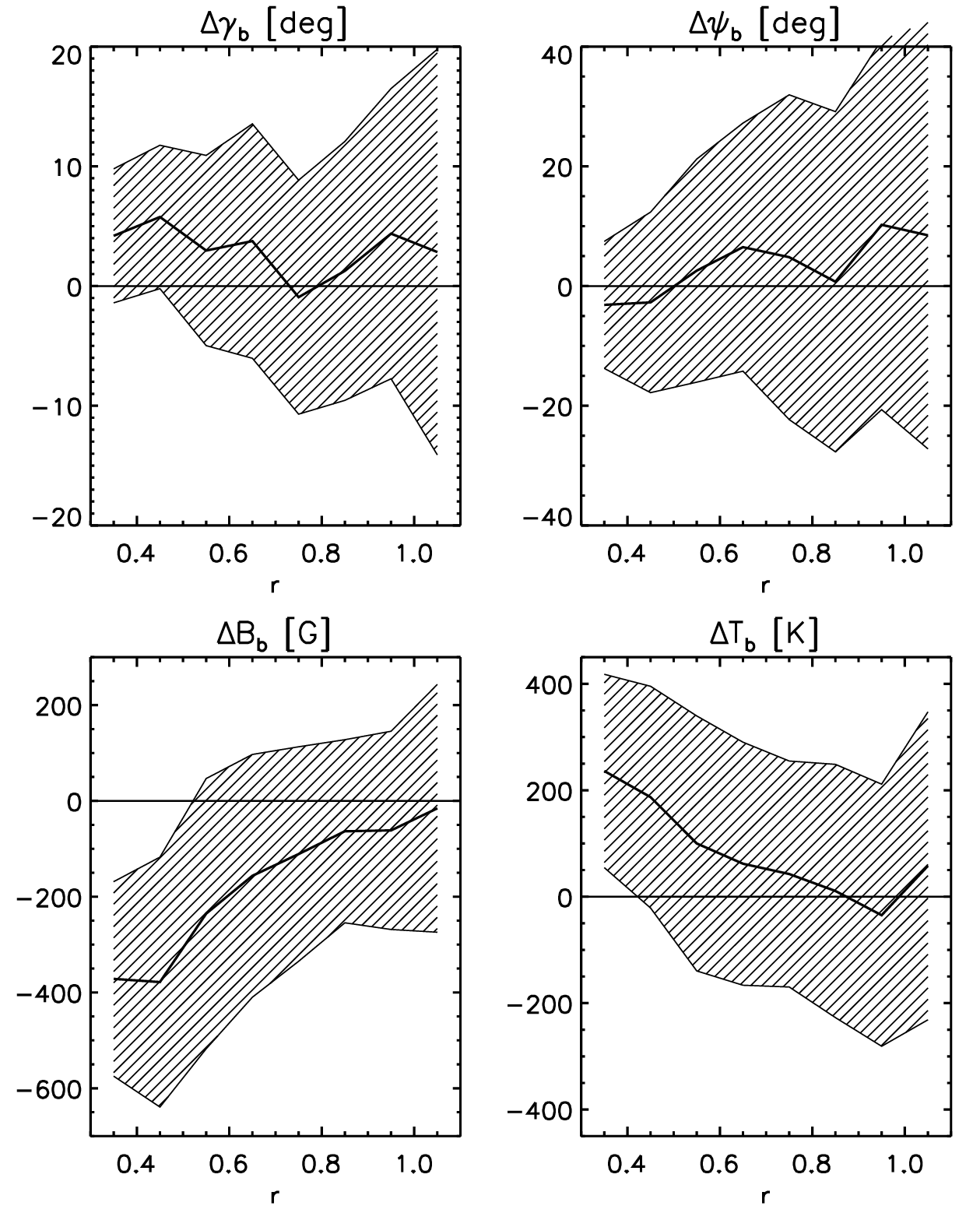}}
\caption{Same as Fig.~\ref{fig:dif_fill_ecs} for the field strength,
  inclination, azimuth, and temperatures at $\log \tau = 0$ of the
  background.\label{fig:dif_back_ecs} \vspace*{-.5em}}
\end{center}
\end{figure}

The mean differences between the filling factors ($\Delta f$) and flow
velocities ($\Delta |v_t|$) of the ECs and the intra-spines are displayed in
Fig.~\ref{fig:dif_fill_ecs} as a function of radial distance. $\Delta f$ is of
the order of 10$\%$, while $\Delta |v_t|$ reaches 0.7~\kms. Similar plots for
the field strength, inclination, azimuth, and temperature differences are 
also presented in Fig.~\ref{fig:dif_fill_ecs}. The values of $\gamma_{\rm t}$ 
and $\psi_{\rm t}$ turn out to be remarkably similar in the ECs and the 
intra-spines. The same happens with $B_{\rm t}$ and $T_{\rm t}$.  This 
implies that the uncombed inversion no longer interprets the ECs as 
perturbations of the magnetic configuration of the intra-spines. 
Figure~\ref{fig:dif_back_ecs} shows the corresponding quantities for 
the background atmosphere: while the passage of ECs is not associated
with changes of $\gamma_{\rm b}$ or $\psi_{\rm b}$, we observe small decreases
of $B_{\rm b}$ and slight increases of $T_{\rm b}$ in the inner penumbra.

\begin{figure}[t]
\begin{center}
\scalebox{0.6}{\includegraphics[bb=74 360 507 700,clip]{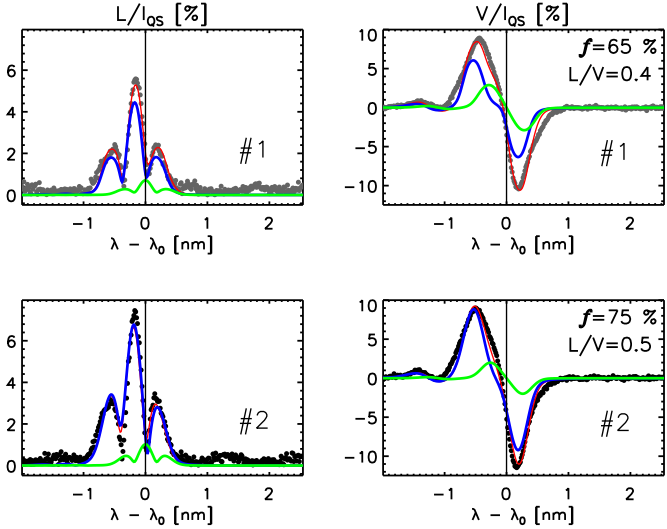}}
\caption{Observed linear and circular polarization profiles of \ion{Fe}{I}
1564.85~nm (dots) emerging from a pixel before (\#1) and during (\#2) the
passage of EC A. The pixel is marked with asterisks and numbers in 
Fig.~\ref{fig:ec_filament_inv2c}. Blue and green lines represent the 
contributions of the tube and background components to the best-fit 
profiles (red lines).\label{fig:prof_ecs_inv2c}}
\end{center}
\end{figure}

To understand why the code returns enhanced filling factors in the ECs,
Fig.~\ref{fig:prof_ecs_inv2c} compares the Stokes profiles emerging from the
same pixel before and during the passage of EC A (cf.\ the numbers marked in
Fig.~\ref{fig:ec_filament_inv2c}). The figure also displays the contribution
of the background and tube atmospheres to the profiles. Note that (a) the EC
passage increases the linear-to-circular polarization ratios mainly through an
enhancement of the linear polarization signal, and (b) the linear polarization
arises almost exclusively from the tube component. This combination of factors
allows the code to explain the EC profiles just by increasing the contribution
of the tube atmosphere, i.e., $f$. In other words: the filling factors of the
ECs are large as a consequence of the high linear-to-circular polarization
ratios they exhibit (Paper~I).


\subsection{Physical origin of the increased filling factors}
\label{sec:origin_ecs}

The large filling factors of the ECs may result from (a) an increase in the
number of penumbral tubes, and/or (b) a greater fraction of the tube's cross
section in the line forming region.

Option (a) seems unlikely because it would require the participation of new
tubes, all of them migrating outward along intra-spines in an organized way as
the ECs. Option (b) would require flux tubes with larger radii or placed higher 
in the photosphere. Both possibilities are ruled out by
Fig.~\ref{fig:pos_width_t}.

Therefore, we deem that perturbations of the thermodynamic properties of the
tubes are more feasible as the source of the filling factor enhancements
associated with the ECs and, consequently, as the origin of the EC
phenomenon. Since no significant temperature variations occur, we suggest that
the EC phenomenon is produced by {\em variations of density/pressure inside
penumbral flux tubes}. This mechanism will be examined in
Sect.~\ref{discussion_uncombed}.

\section{Discussion}
\label{discussion}
\subsection{One-component interpretation}
In Paper I we found that ECs are penumbral structures characterized by larger
Doppler shifts, stronger linear polarization signals, and larger Stokes $V$
area asymmetries than both spines and intra-spines. We suggested that these
were observational signatures of stronger Evershed flows and more horizontal
magnetic fields in the ECs as compared with the rest of the penumbra. Our
one-component inversions seem to confirm this picture.

\begin{figure*}
\begin{center}
\scalebox{0.33}{\includegraphics[bb=60 -13 552 808,clip]{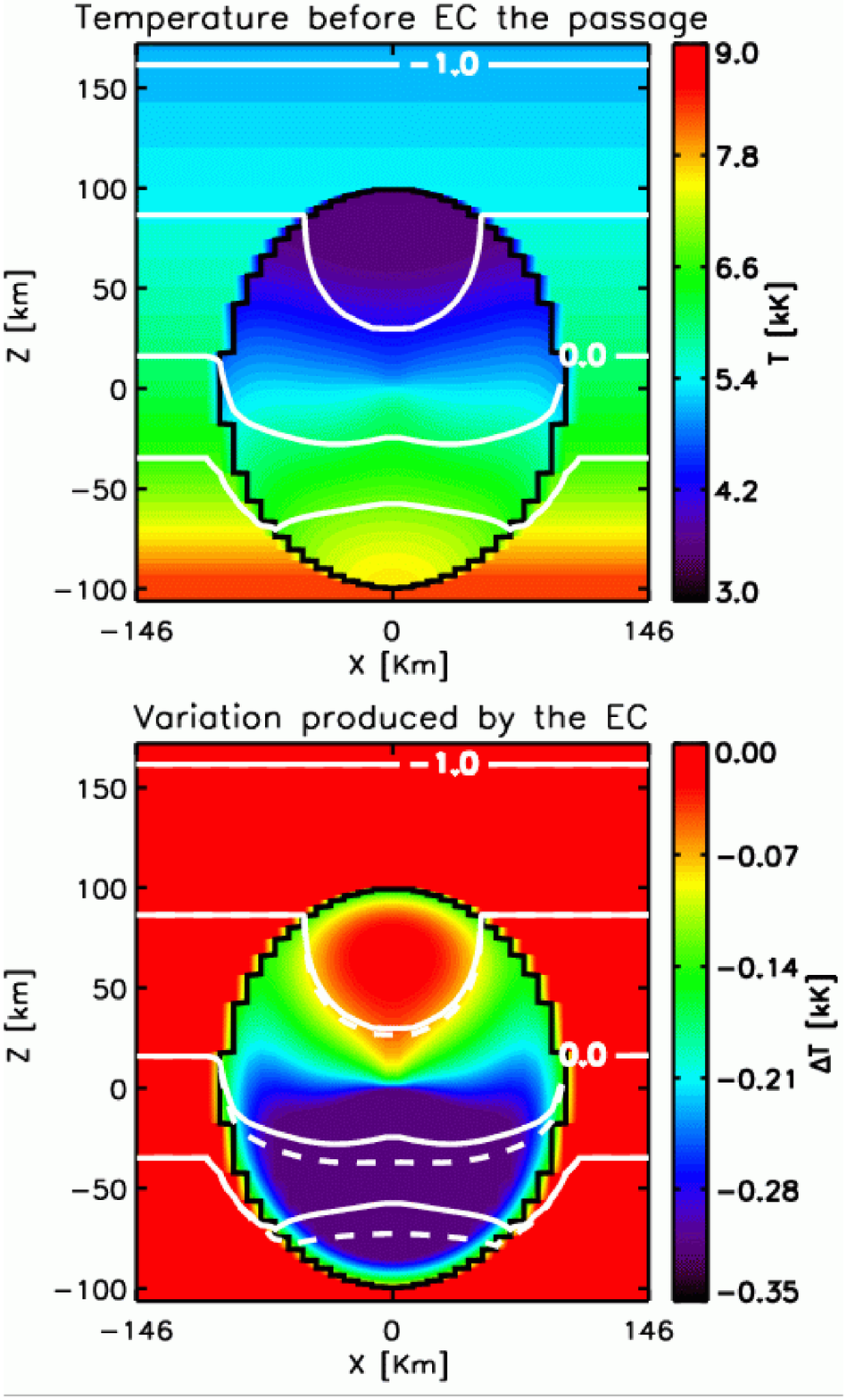}}
\scalebox{0.33}{\includegraphics[bb=60 -13 552 808,clip]{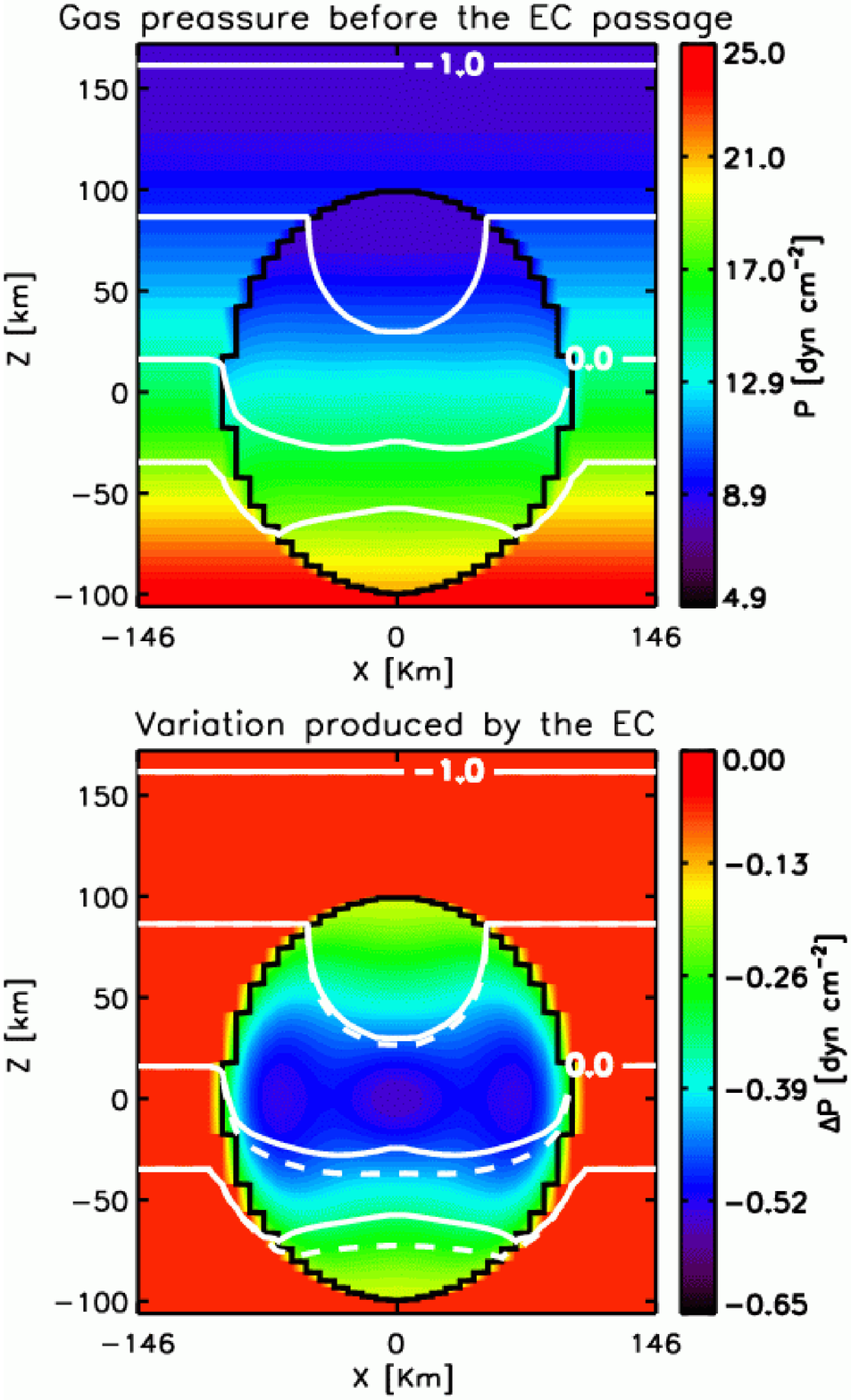}}
\scalebox{0.33}{\includegraphics[bb=60 -13 552 808,clip]{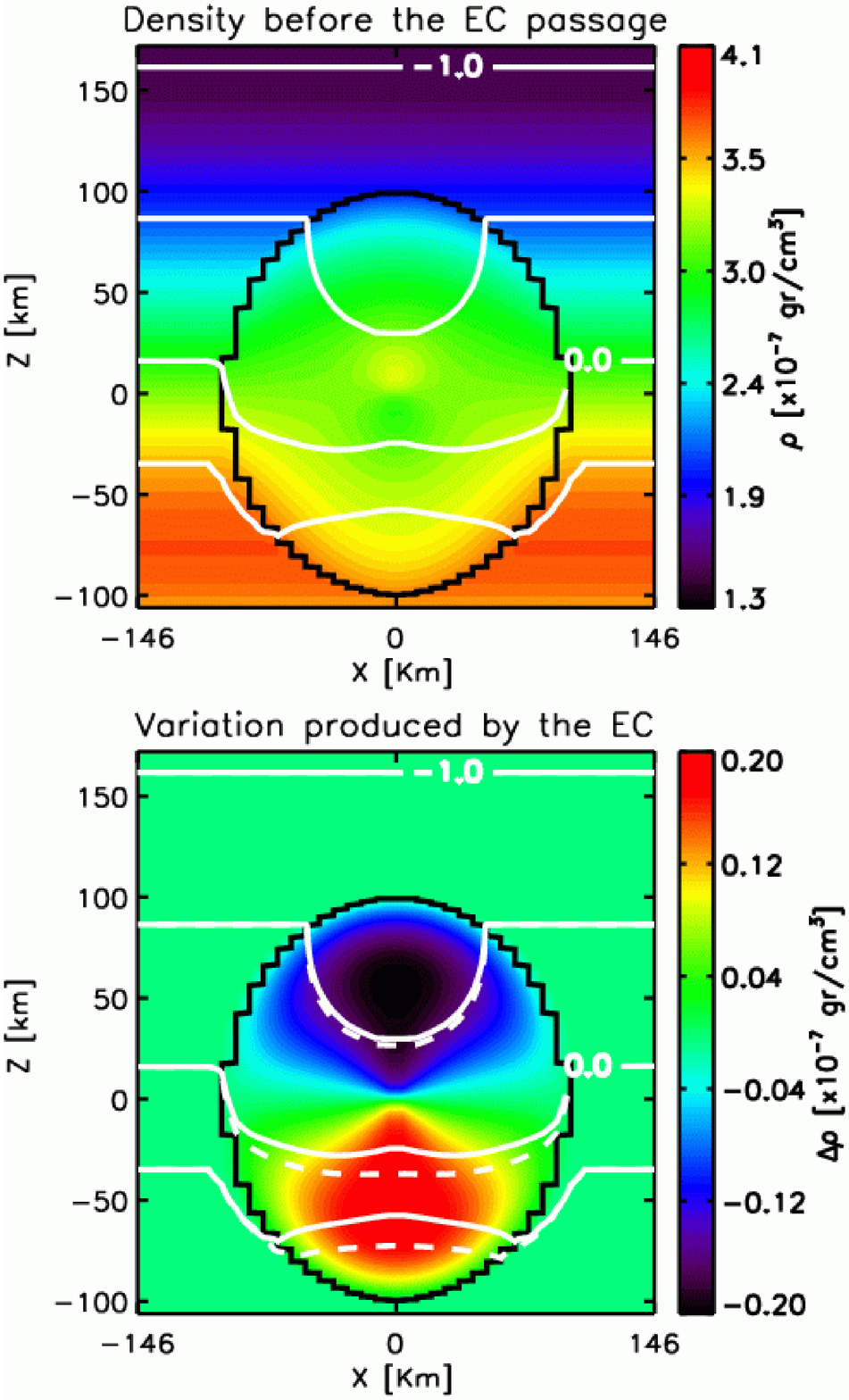}}
\caption{{\em Top panels:} Temperature, gas pressure, and density in
the tube and their surroundings simulating the conditions of the atmosphere 
before the EC passage. {\em Bottom panels:} Variation of the temperature, 
gas pressure, and density produced by the EC. The tube is enclosed by the 
black lines. Solid and dashed white lines indicate isocontours of optical 
depth ($\log \tau = -1, -0.5, 0, 0.5$) before and during the EC 
passage, respectively.
\label{fig:models}}
\end{center}
\end{figure*}

Two different mechanisms have been proposed to explain the origin of ECs:
magnetoacoustic waves superimposed on a steady or quasi-steady flow
\citep{2003ApJ...584..509G}, and $\Omega$ or $\mho$ kinks propagating along
field lines \citep{1998ApJ...492..402R, 2003A&A...397..757R}. Another
possibility is that ECs are the signatures of sea-serpent field lines
associated with moving penumbral tubes \citep{2002AN....323..303S}. None of
the three mechanisms seems to be compatible with the magnetic field geometry
deduced from the inversions. The reason is that they should produce both an
increase and a decrease of the field inclination in the ECs, but the former 
is not detected.  

We cannot exclude that higher resolution observations may solve the problem in
the future. With the present data, however, the conclusion is that origin of
the magnetic field perturbations indicated by the one-component inversions
remains unknown.

\subsection{Uncombed interpretation}
\label{discussion_uncombed}

We have carried out simple numerical experiments to examine whether or not
larger linear-to-circular polarization ratios and, consequently, an
enhancement of $f$, may result from density and pressure variations inside
penumbral flux tubes. To that end we use the model of flux tubes in mechanical
equilibrium proposed by \citet{2007.borrero}. In this model the tubes are not
thin, i.e., the physical properties change over their cross sections.

We assume two different equilibrium configurations to model the conditions of
the tube and background atmospheres in intra-spines before and during the
passage of an EC. In order to emulate the conditions \emph{before} the EC
passage we consider a horizontal ($\gamma_{\rm t} = 90^\circ$) tube with
a field strength of 1000~G that channels an Evershed flow of $|v_{\rm t}|
=4.5$~\kms.  The tube is located at the heliocentric angle of the observations
(43\degree~on 30 June) with its axis pointing along the line of symmetry. It
is surrounded by a background field with $\gamma_{\rm b} = 60^\circ$ and
$B_{\rm b} = 450$~G.  The upper panels of Fig.~\ref{fig:models} show the
temperatures, gas pressures, and densities derived from the model. The
$x$-axis is perpendicular to the tube axis and the $z$-axis represents the
line of sight.

The only property that seems to change \emph{during} the passage of an EC is
the strength of the background field. Thus, we model the new situation using
the same parameters for the tube component ($\gamma_{\rm t} = 90^\circ$,
$B_{\rm t} = 1000$~G, $\gamma_{\rm b} = 60^\circ$) but a weaker field of
$B_{\rm b} = 350$~G in the background atmosphere. To maintain lateral
pressure balance, the gas pressure and density of the tube decrease 
in response to the weakening of the background field (bottom panels of
Fig.~\ref{fig:models}).

The pressure and density variations associated with the EC move the optical
depth scale toward deeper layers, which increases the fraction of the tube
inside the line formation region (compare the solid and dashed lines in the
bottom panels of Fig.~\ref{fig:models}). It is important to remark that the
variation of the optical depth scale is negligible in the upper half of the
tube, becoming significant only in the lower half. This may explain why the
position of the tube's upper boundary is not seen to change with the EC
passage.

\begin{figure}
\centering
\scalebox{0.6}{\includegraphics[bb=74 360 507 700,clip]{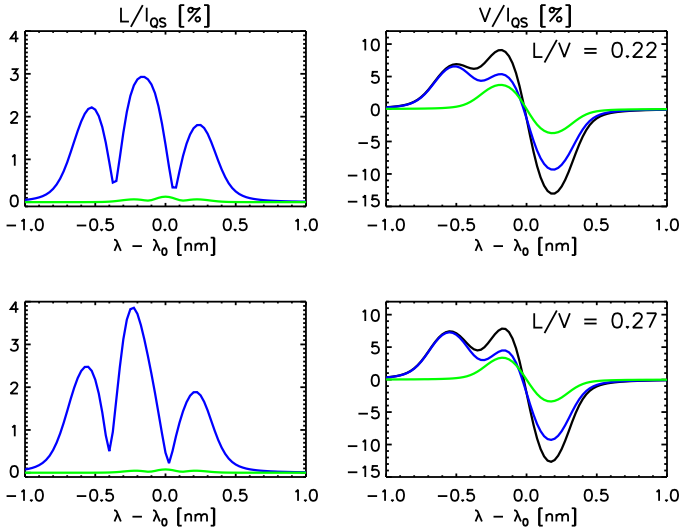}}
\caption{Synthetic linear and circular polarization profiles of \ion{Fe}{i}
  1564.8~nm (black curves) simulating the conditions before ({\em top}) and
  during ({\em bottom}) the EC passage. Blue and green profiles represent the
  contributions of the tube and background atmospheres to the emergent
  profiles, respectively.\label{fig:prof_ir}}
\end{figure}

The downward shift of the optical depth scale increases the observed $L/V$
ratios. This is demonstrated in Fig.~\ref{fig:prof_ir}, where we show the
total linear and circular polarization profiles of \ion{Fe}{i} 1565~nm before
and during the EC passage. The contributions of the tube and background
atmospheres to the emergent profiles are indicated with blue and green
lines, respectively. When the EC moves along the intra-spine, the
linear-to-circular polarization ratio grows from $L/V=0.22$ to
$L/V=0.27$. Since the amount of circular polarization remains roughly the
same, the increase in $L/V$ is due to an increase in the linear polarization,
much in the same way as observed (Fig.~\ref{fig:prof_ecs_inv2c}). The larger
$L/V$ ratio is mainly due to a larger flux tube contribution. Since the
magnetic properties of the tube are the same before and during the EC passage,
the only possible cause is a change in the thermodynamic properties 
of the tube: {\em a decrease in the gas pressure and
density}\footnote{To confirm this point, we have repeated the experiment
assuming that the gas pressure in the tube increases during the passage of the
EC. This is induced through a larger background magnetic field ($B_{\rm b} =
550$~G vs $B_{\rm b} = 450$~G).  As expected, the larger gas pressures and
densities in the tube shift the optical depth scale toward higher layers,
reducing the contribution of the tube to the linear polarization profiles and,
thus, the $L/V$ ratio.}.  The larger flux tube contribution is interpreted by
uncombed inversions as enhancements of the filling factor.

Our experiments support the idea that the $L/V$ increase observed during the
passage of ECs is the result of lower gas pressures and densities in 
the penumbral tubes that carry the Evershed flow. With the model of
\citet{2007.borrero}, these conditions are forced imposing a weaker external
magnetic field. Such a field is consistent with the observations
(Fig.~\ref{fig:dif_back_ecs}), but its origin remains unclear.

\section{Summary}
\label{summary}

In this paper we have interpreted the Stokes profiles emerging from Evershed
clouds (ECs) using simple one-component and more sophisticated uncombed
inversions of 4 visible and 3 infrared lines. The inversions have allowed us
to determine the physical properties of ECs, the ultimate goal being to
understand the nature of the EC phenomenon.

The one-component inversions confirm many of the properties indicated by the
line parameter study of Paper I. The results of these inversions suggest that
ECs are structures having more inclined fields and stronger Evershed flows
than the rest of the penumbra. ECs move along intra-spines toward the outer
boundary of the spot. In their journey, they change the magnetic configuration
of the intra-spines, which only recover their initial state after the passage
of the ECs. Thus we suggest that the EC phenomenon is caused by a perturbation
of the magnetic field of the intra-spines. None of the scenarios proposed so
far for the EC phenomenon seem to be capable of explaining these
findings. Both magnetoacoustic waves and magnetic kinks should give rise to
vector fields pointing to the solar interior, but we do not detect them.

\begin{figure}
\centering
\scalebox{0.3}{\includegraphics{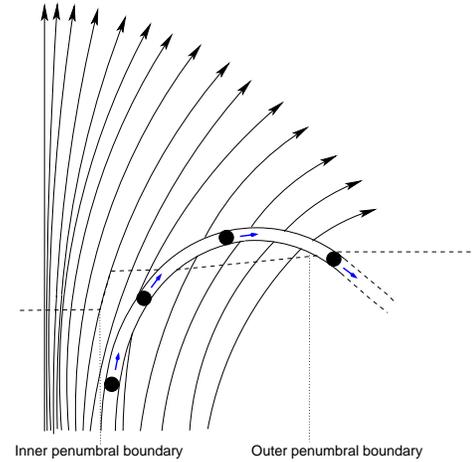}}
\caption{Cartoon of a type I EC propagating along a penumbral flux tube which
dives back to the solar interior at the outer sunspot edge. The EC is
represented by the black circles. The background field is indicated by the
arrows. The dashed line marks the $\tau=1$ level, and the vertical dotted
lines the penumbral boundaries. For visualization purposes, the spot 
is significantly compressed in the horizontal direction. This makes the 
flux tube appear to be elevated, but in reality it is a deep-lying 
tube that never reaches high photospheric layers.
\label{cartoon}}
\end{figure}

When we account for the fine structure of the penumbra, the EC phenomenon is
no longer interpreted as a real perturbation of the magnetic field of the
intra-spines. The uncombed inversions suggest that the physical properties of
the penumbral flux tubes are rather similar in the ECs and the intra-spines
hosting them. The background atmospheres are also similar, except for a small
decrease in the field strength and slightly larger temperatures.  The only
significant difference observed in the intra-spines during the passage of an
EC is an enhancement of the filling factor of the tube component, i.e., a
larger visibility of the flux tubes that carry the Evershed flow.  The
inversions indicate that the enhanced visibility is not due to changes in
the position or radius of the tubes.  We therefore propose that the EC 
phenomenon is the result of variations of gas pressure and density in the 
flux tubes. Such variations would propagate toward the outer sunspot boundary, 
producing the motion of the ECs.  Our interpretation is summarized in
Fig.~\ref{cartoon}.  Type I ECs vanish at the edge of the penumbra because the
tubes return to the solar interior and the density/pressure perturbations
go out of the line forming region. Type II ECs disappear in the sunspot moat
(cf.\ Paper I), so the flux tubes associated with them have to continue well
beyond the outer penumbral border \citep[see][]{cabrera.2008}.

We have performed simple numerical experiments to examine the feasibility of
this scenario.  We find that a decrease of the gas pressure in the upper half
of the tube implies a reduction of the density, which shifts the optical depth
scale toward deeper atmospheric layers. The shift does not change the optical
depth of the tube's upper boundary, but makes the tube occupy a larger
fraction of the line forming region. The result is an increase in the
linear-to-circular polarization ratios associated with the ECs, which are
interpreted by the inversion code as larger filling factors.  This mechanism
holds promise to explain the EC phenomenon. However, further numerical and
theoretical work is needed to understand the origin of the pressure/density
perturbations.  An interesting possibility is that they are the signatures of
shocks inside penumbral flux tubes, like the ones detected by
\cite{2005A&A...436..333B}.

\begin{acknowledgements}
We thank P.J. Guti\'errez and D.\ Orozco Su\'arez for sharing their computing
time with us. C.\ Westendorp Plaza generously offered his routines to plot
vector fields in 3D. This work has been supported by the Spanish MEC under
project ESP2006-13030-C06-02 and Programa Ram\'on y Cajal. The German VTT is
operated by the Kiepenheuer-Institut f\"ur Sonnenphysik on the Observatorio
del Teide of the Instituto de Astrof\'{\i}sica de Canarias. The DOT is
operated by Utrecht University at the Spanish Observatorio del Roque de los
Muchachos, also of the Instituto de Astrof\'{\i}sica de Canarias.
\end{acknowledgements}

\bibliographystyle{aa}

\end{document}